\documentclass[prd,aps,10pt,nofootinbib,twocolumn,superscriptaddress,preprintnumbers,balancelastpage,longbibliography]{revtex4-1}

\usepackage{amsmath,amssymb}	
\usepackage{mathtools}
\usepackage{hyperref}
\usepackage[dvipsnames]{xcolor}
\usepackage{slashed}
\usepackage{xspace}
\usepackage{comment}
\usepackage{fancyhdr}
\usepackage{braket}
\usepackage{graphicx}
\usepackage{float}
\usepackage{siunitx}
\usepackage{physics}

\usepackage{lipsum}
\usepackage{booktabs, tabularx, makecell, multirow, xspace}
\usepackage{listings}

\renewcommand{\arraystretch}{1.8}
\newcolumntype{C}[1]{>{\centering\let\newline\\\arraybackslash\hspace{0pt}}m{#1}}

\definecolor{linkcolor}{rgb}{0.7752941176470588, 0.22078431372549023, 0.2262745098039215}

\hypersetup{colorlinks=true,
linkcolor=linkcolor,
citecolor=linkcolor,
urlcolor=linkcolor,
linktocpage=true,
pdfproducer=medialab,
}

\newcommand{\Eq}[1]{Eq.~(\ref{#1})}

\definecolor{nicered}{rgb}{0.7,0.1,0.1}
\definecolor{nicegreen}{rgb}{0.1,0.5,0.1}
\newcommand{\Neff}{$N_\text{eff}$}

\DeclareSIUnit\electronvolt{e\kern-.05em V}

\newcommand\Tstrut{\rule{0pt}{2.6ex}}         
\newcommand\Bstrut{\rule[-0.9ex]{0pt}{0pt}}   

\definecolor{codegreen}{rgb}{0,0.6,0}
\definecolor{codegray}{rgb}{0.5,0.5,0.5}
\definecolor{codepurple}{rgb}{0.58,0,0.82}
\definecolor{backcolour}{rgb}{0.95,0.95,0.92}

\lstdefinestyle{pystyle}{
    language = Python,  
    commentstyle=\color{codegreen},
    keywordstyle=\color[rgb]{0,0,0.75},
    keywordstyle=[1]\color[rgb]{0,0,0.75},
    keywordstyle=[2]\color[rgb]{0,0,0.75},
    keywordstyle=[3]\color[rgb]{0,0,0.75},
    keywordstyle=[4]\color[rgb]{0,0,0.75},
    commentstyle=\color[rgb]{0.133,0.545,0.133},
    stringstyle=\color{codepurple},
    basicstyle=\ttfamily,
    breakatwhitespace=false,         
    breaklines=true,                 
    captionpos=b,                    
    keepspaces=false,                 
    numbers=none,                    
    numbersep=5pt,                  
    showspaces=false,                
    showstringspaces=false,
    showtabs=false,                  
    tabsize=2,
    morekeywords={True, False, len},
    columns=flexible
}

\lstdefinestyle{mathstyle}{
    language = Mathematica,  
    commentstyle=\color{codegreen},
    keywordstyle=\color[rgb]{0,0,0.75},
    keywordstyle=[1]\color[rgb]{0,0,0.75},
    keywordstyle=[2]\color[rgb]{0,0,0.75},
    keywordstyle=[3]\color[rgb]{0,0,0.75},
    keywordstyle=[4]\color[rgb]{0,0,0.75},
    commentstyle=\color[rgb]{0.133,0.545,0.133},
    stringstyle=\color{codepurple},
    basicstyle={\ttfamily \small},
    breakatwhitespace=false,         
    breaklines=true,                 
    captionpos=b,                    
    keepspaces=false,                 
    numbers=none,                    
    numbersep=5pt,                  
    showspaces=false,                
    showstringspaces=false,
    showtabs=false,                  
    tabsize=2,
    morekeywords={True, False, len},
    columns=flexible
}

\makeatletter 
    
\renewcommand\onecolumngrid{
\do@columngrid{one}{\@ne}
\def\set@footnotewidth{\onecolumngrid}
\def\footnoterule{\kern-6pt\hrule width 1.5in\kern6pt}
}

\renewcommand\twocolumngrid{
        \def\footnoterule{
        \dimen@\skip\footins\divide\dimen@\thr@@
        \kern-\dimen@\hrule width.5in\kern\dimen@}
        \do@columngrid{mlt}{\tw@}
}

\makeatother    

\begin{document}

\title{Joint CMB and BBN Constraints on Light Dark Sectors with Dark Radiation}

\author{Cara Giovanetti}
\email{cg3566@nyu.edu}
\thanks{ORCID: \href{https://orcid.org/0000-0003-1611-3379}{0000-0003-1611-3379}}
\affiliation{Center for Cosmology and Particle Physics, Department of Physics, New York University, New York, NY 10003, USA}

\author{Mariangela Lisanti}
\email{mlisanti@princeton.edu}
\thanks{ORCID: \href{https://orcid.org/0000-0002-8495-8659}{0000-0002-8495-8659}}
\affiliation{Department of Physics, Princeton University, Princeton, NJ 08544, USA}
\affiliation{Center for Computational Astrophysics, Flatiron Institute, 162 Fifth Ave, New York, NY 10010, USA}

\author{Hongwan Liu}
\email{hongwanl@princeton.edu}
\thanks{ORCID: \href{https://orcid.org/0000-0003-2486-0681}{0000-0003-2486-0681}}
\affiliation{Center for Cosmology and Particle Physics, Department of Physics, New York University, New York, NY 10003, USA}
\affiliation{Department of Physics, Princeton University, Princeton, NJ 08544, USA}

\author{Joshua T. Ruderman}
\email{ruderman@nyu.edu}
\thanks{ORCID: \href{https://orcid.org/0000-0001-6051-9216}{0000-0001-6051-9216}}
\affiliation{Center for Cosmology and Particle Physics, Department of Physics, New York University, New York, NY 10003, USA}

\date{\protect\today}

\begin{abstract}
Dark sectors provide a compelling theoretical framework for thermally producing sub-GeV dark matter, and motivate an expansive new accelerator and direct-detection experimental program.  We demonstrate the power of constraining such dark sectors using the measured effective number of neutrino species, $N_\text{eff}$, from the Cosmic Microwave Background~(CMB) and primordial elemental abundances from Big Bang Nucleosynthesis~(BBN).  As a concrete example, we consider a dark matter particle of arbitrary spin that interacts with the Standard Model via a massive dark photon, accounting for an arbitrary number of light degrees of freedom in the dark sector.  We exclude dark matter masses below $\sim 4$~MeV at 95\% confidence for all dark matter spins and dark photon masses.  These bounds hold regardless of additional new light, inert degrees of freedom in the dark sector, and for dark matter-electron scattering cross sections many orders of magnitude below current experimental constraints.  The strength of these constraints will only continue to improve with future CMB experiments.
\end{abstract}

\maketitle

\noindent
\noindent\textbf{Introduction.---}The exquisite precision of Cosmic Microwave Background~(CMB) and Big Bang Nucleosynthesis~(BBN) measurements has historically played an important role in constraining the properties of dark matter (DM)~\cite{Kolb_1986, Sarkar:1995dd, Serpico:2004nm,Iocco:2008va, Pospelov:2010hj, Ho:2012ug, Boehm:2013jpa, Nollett:2013pwa, Nollett_2014b,Wilkinson:2016gsy,Green:2017ybv,Escudero_2019,Depta:2019lbe,Sabti:2019mhn,sabti2021implications}. The introduction of new particles in the dark sector can affect the expansion rate of the Universe, as well as the temperature of Standard Model~(SM) particles, thereby leaving distinctive signatures on both elemental abundances and the effective number of neutrino species, \Neff. In this Letter, we demonstrate how to compute joint CMB and BBN constraints for generic dark sectors, using the example of a sub-GeV DM species accompanied by a massive dark photon and an arbitrary number of light, inert degrees of freedom.

Joint CMB and BBN constraints have been obtained for a single DM particle in thermal equilibrium with the SM at early times~\cite{Boehm:2012gr,Boehm:2013jpa,Steigman:2013yua,Nollett:2013pwa,Nollett_2014b,Wilkinson:2016gsy,Depta:2019lbe,Sabti:2019mhn,sabti2021implications}.  For example, Refs.~\cite{Sabti:2019mhn,sabti2021implications} find that an electromagnetically coupled DM particle must have mass $m_\chi \gtrsim \SI{5}{\mega\eV}$ at 95\% confidence, depending on its spin. However, the need for joint constraints is more significant for dark sectors. This was underscored by Refs.~\cite{Steigman:2013yua,Nollett:2013pwa,Green:2017ybv}, which obtained joint CMB and BBN constraints for electromagnetically coupled DM accompanied by additional relativistic degrees of freedom in the dark sector. In this model, CMB-only constraints cannot break the degeneracy between DM entropy injection---which heats photons relative to neutrinos after neutrino decoupling, leading to a lower value of \Neff---and new inert, relativistic degrees of freedom.  Primordial elemental abundances are affected in different ways by the radiation energy density and the neutrino temperature during BBN, and can therefore break this degeneracy.  

Beyond these simple models, CMB and BBN constraints have the potential to play an important role in our understanding of well-motivated dark sectors, many of which yield viable thermal relics in the keV--GeV mass range (see, e.g., Refs.~\cite{Boehm:2003hm, Pospelov:2007mp, Feng:2008ya,Kaplan:2009ag,Cohen:2010kn, Chu:2011be,Agashe:2014yua,Hochberg:2014dra,Hochberg:2014kqa,Lee:2015gsa,Hochberg:2015vrg, Izaguirre:2015yja,Bernal:2015xba,Kopp:2016yji,Choi:2016tkj,Dey:2016qgf,Bernal:2017mqb,Cline:2017tka, Knapen:2017xzo}). These dark sectors commonly have multiple states that interact with the SM through portal interactions, which need to be properly accounted for when determining the joint CMB and BBN constraints. Furthermore, new numerical methods~\cite{Escudero:2020dfa, Pitrou:2018cgg} now allow for such joint constraints to be calculated with the inclusion of many potentially important effects, including noninstantaneous neutrino decoupling and BBN nuclear rate uncertainties. 

We focus on a scenario where the DM particle $\chi$ couples to a massive $U(1)'$ dark photon $A'$ that is  kinetically mixed with the SM photon. We also include the possibility of new inert, relativistic degrees of freedom, which has been used to avoid CMB-only \Neff\, constraints due to the aforementioned degeneracy with DM entropy injection (see, e.g., Refs.~\cite{Boehm:2013jpa,Essig:2015cda,Wilkinson:2016gsy,Izaguirre:2017bqb,Berlin:2018bsc}).  This model is one of the standard benchmarks for the nascent experimental program for the direct detection of DM-electron scattering~\cite{Battaglieri:2017aum}.  

Using the 2018 Planck results~\cite{Aghanim:2018eyx}, as well as the primordial elemental abundances from Refs.~\cite{Zyla:2020zbs, Cooke:2017cwo}, we robustly constrain the DM mass $m_\chi$ as a function of its spin and the dark photon mass $m_{A'}$.  For example, when $m_{A'}/m_\chi = 3$, we exclude complex scalar~(Dirac fermion) DM below $m_\chi \sim \SI{5.2}{\mega\eV}$~(\SI{7.9}{\mega\eV}) at 95\% confidence for DM-electron scattering cross sections that are many orders of magnitude below current constraints.  These results apply regardless of the number of inert, relativistic degrees of freedom in the model, thereby circumventing a key weakness of previous cosmological constraints of this kind.  They will also strengthen with future CMB measurements.

\vspace{0.1in}
\noindent\textbf{Methodology.---}We compute the effect of a dark sector model with parameters $\boldsymbol\theta$ on the effective number of relativistic degrees of freedom at late times, $N_{\rm eff}$.  For our benchmark model, 
\begin{equation}
    N_{\textrm{eff}}= 3\left[\frac{11}{4}\left(\frac{T_{\nu}}{T_{\gamma}}\right)^3_0\right]^{4/3}\left(1+\frac{\Delta N_{\nu}}{3}\right) \, .
    \label{eq:Neff}
\end{equation}
Here, $T_\gamma$ and $T_\nu$ are the photon and neutrino temperatures, respectively, and $\Delta N_{\nu}$ is the ratio of the energy density $\rho_{\xi}$ of inert, relativistic degrees of freedom $\xi$ in the dark sector to that of a single neutrino at late times, taking $\rho_\xi \propto a^{-4}$ throughout cosmic history.  The benchmark model consists of two free parameters, $m_\chi$ and $\Delta N_\nu$; $m_{A'}$ is set as a constant multiple of $m_\chi$.   The subscript `0' denotes a late point in time when $T_\gamma \ll m_\chi$, $m_{A'}$, and $m_e$, the electron mass.  As demonstrated by \Eq{eq:Neff}, annihilations that preferentially inject entropy into the photon bath decrease \Neff\, relative to the standard cosmological value.  This decrease, which depends on $m_\chi$, can be compensated for by increasing $\Delta N_\nu$.  Appendix A reviews how dark sectors impact $T_\gamma$ and $T_\nu$. 

We also determine the effect of the dark sector on the primordial abundances of elements after BBN has ended.  We only consider Y$_{\textrm{P}}$ and D/H, the ratio of the abundance by mass of helium-4 and the ratio of the abundance of deuterium to hydrogen, respectively; extending the analysis to other elements is straightforward. We compute Y$_{\textrm{P}}$ and D/H for a given dark sector model over the range $\Omega_b h^2 \in [0.0218,0.0226]$---much broader than the Planck uncertainty on this parameter~\cite{Aghanim:2018eyx}---since the production of light elements is highly sensitive to the baryon-to-photon ratio~\cite{Nollett:2013pwa}.  

These calculations were performed using the public codes \texttt{nudec\_BSM}~\cite{Escudero_2019,Escudero:2020dfa} and \texttt{PRIMAT}\footnote{We use the latest version of \texttt{PRIMAT}~\cite{Pitrou:2020etk}, which includes the recently updated measurement of the D + p $\rightarrow$ $^3$He + $\gamma$ cross section~\cite{Mossa:2020gjc}.}~\cite{Pitrou:2018cgg,Pitrou:2020etk}, which we modify to include a DM particle (of arbitrary spin), a dark photon, and $\Delta N_\nu$.  Our modified \texttt{nudec\_BSM} first computes $T_\gamma$ and $T_\nu$, as well as the Hubble rate and scale factor as functions of time, self-consistently including the effects of non-instantaneous neutrino decoupling and QED corrections~\cite{Escudero_2019,Escudero:2020dfa}. The \texttt{nudec\_BSM} output is then used by our modified version of \texttt{PRIMAT} to obtain an accurate prediction of the elemental abundances. This method can be easily extended to any arbitrary dark sector. 

To assess the consistency of the computed $N_\text{eff}(\boldsymbol\theta)$, $\textrm{Y}_{\textrm{P}}(\boldsymbol\theta, \Omega_b h^2)$, and D/H$(\boldsymbol\theta, \Omega_b h^2)$ with CMB and BBN measurements, we perform a hypothesis test on our model parameters by constructing a profile likelihood ratio. Explicitly, we define
\begin{alignat}{1}
    L(\boldsymbol\theta, \Omega_b h^2) = L_\text{BBN}(\boldsymbol\theta, \Omega_b h^2) L_\text{CMB}(\boldsymbol\theta, \Omega_b h^2) \,,
\end{alignat}
where $L$ is the Gaussian likelihood of the parameters $\boldsymbol{\theta}$ and $\Omega_b h^2$. The contribution from the CMB measurements, $L_\text{CMB}$, is computed using the central values and covariance matrix from Planck for a fit with the six $\Lambda$CDM parameters, plus \Neff\ and Y$_\text{P}$~\cite{Aghanim:2018eyx,Planck_Legacy_Archive}, with 
\begin{equation}
    N_\text{eff} = 2.926 \pm 0.286 \, .
\end{equation}
For comparison, we also show projected results for the upcoming Simons Observatory~\cite{SimonsObservatory:2018koc}.

\begin{figure}[t]
    \centering
    \includegraphics[width=0.49\textwidth]{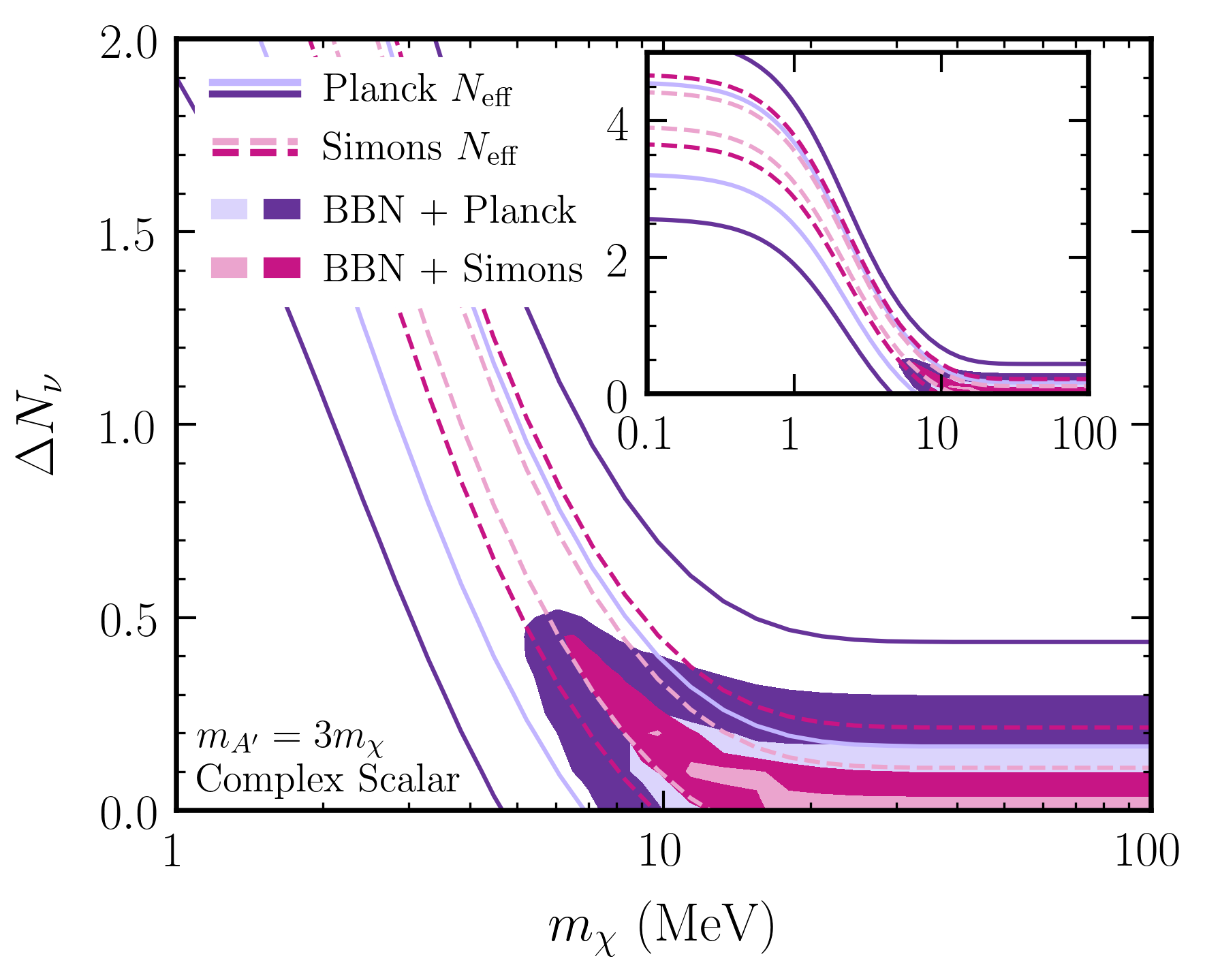}
    \caption{The allowed region of dark matter mass $m_\chi$ and $\Delta N_\nu$, for a complex scalar particle and a dark photon with mass $3m_{\chi}$.  The inset shows the same contours over a larger range of $m_{\chi}$ and $\Delta N_{\nu}$.  The purple (pink) lines denote the Planck (projected Simons Observatory) $N_{\rm eff}$ 68/95\% confidence regions.  The corresponding shaded regions correspond to the bounds including BBN measurements. The BBN measurements clearly play a powerful role in setting the lower limit on the dark matter mass, even when $\Delta N_\nu > 0$ is allowed. The value of $\Omega_b h^2$ that maximizes the likelihood was chosen for each parameter point.}
    \label{fig:Joint_Constraint_Blobs}
\end{figure}

\begin{figure*}[t]
    \centering
    \includegraphics[width=0.47\textwidth]{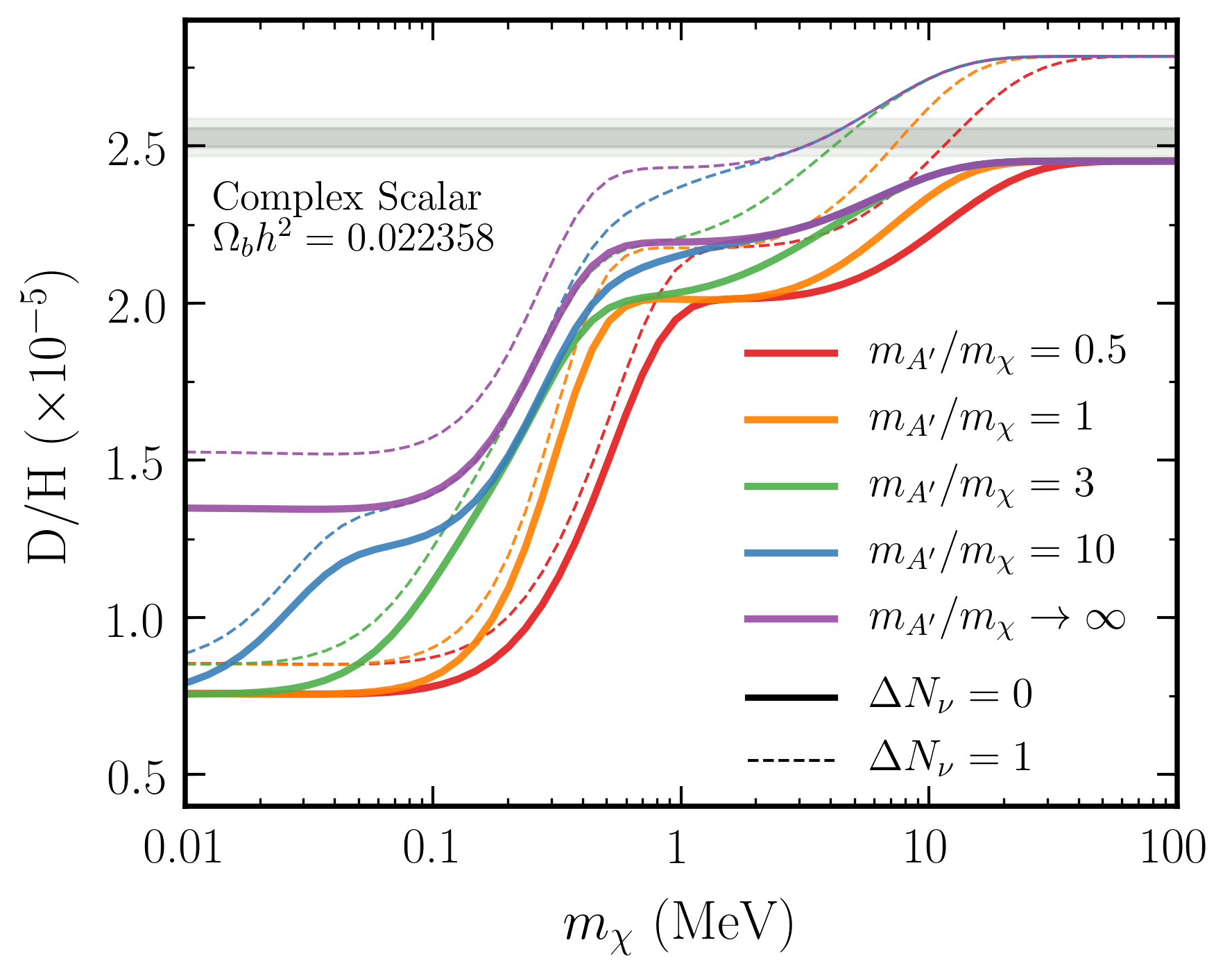}
    \includegraphics[width=0.47\textwidth]{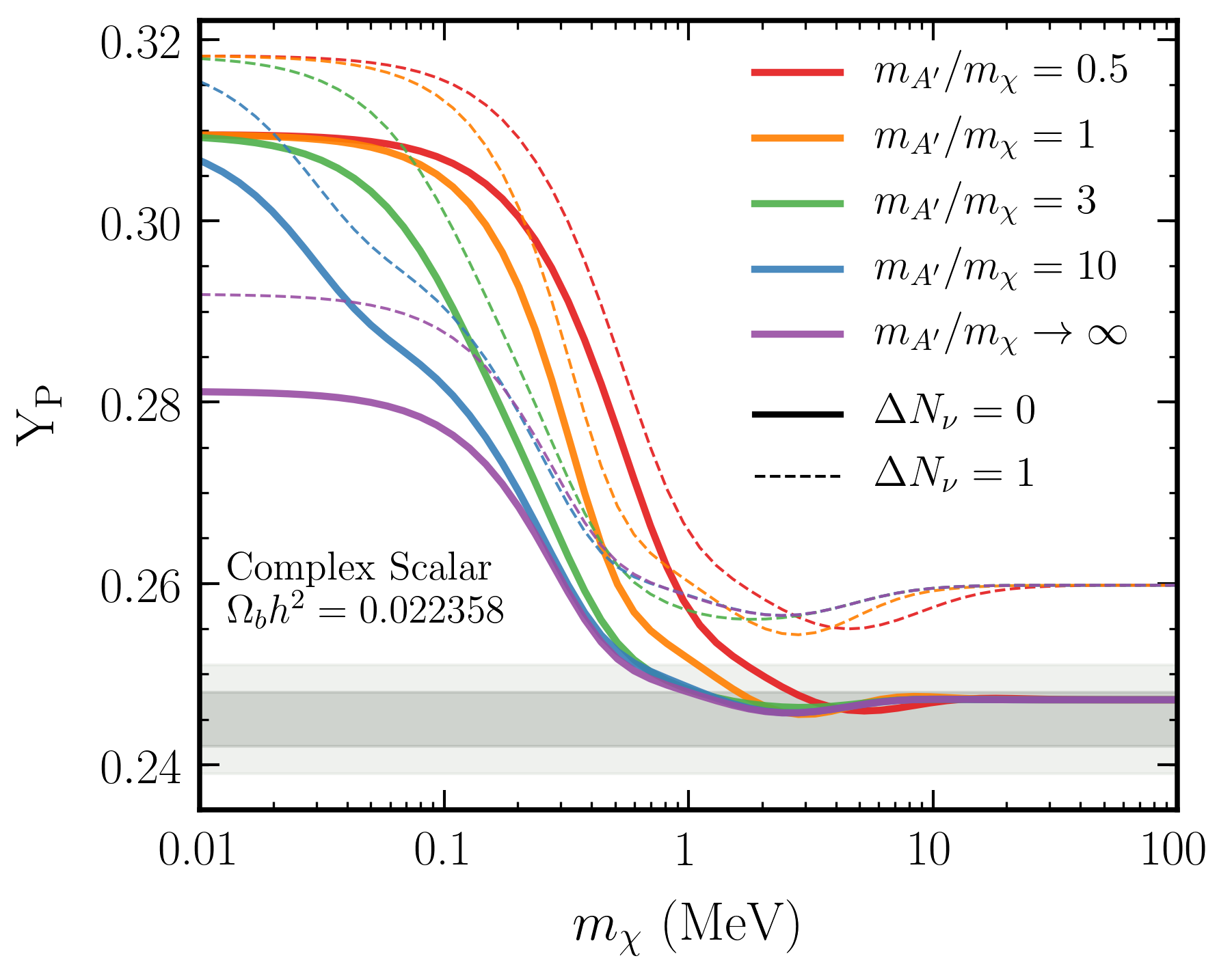}
    \caption{Predictions for D/H (left) and Y$_{\textrm{P}}$ (right) for a complex scalar particle $\chi$ of mass $m_\chi$.  The solid (dashed) lines correspond to $\Delta N_{\nu}=0~(1)$.  The results are shown for five different ratios of dark photon to DM mass: $m_{A'}/m_\chi = 0.5, 1, 3, 10$, and approaching $\infty$ in red, orange, green, blue, and purple, respectively.     
    The 1 and 2$\sigma$ uncertainties on the measured values of D/H~\cite{Cooke:2017cwo} and Y$_{\textrm{P}}$~\cite{Zyla:2020zbs} are indicated by the gray horizontal bands.  The standard cosmology predictions for D/H and Y$_{\rm P}$ are consistent with the results in Ref.~\cite{Pitrou:2020etk}.  $\Omega_b h^2$ is fixed to $0.022358$.}
    \label{fig:BBN_only}
\end{figure*}

The contribution from BBN, $L_\text{BBN}$, is computed using the central values of the observed elemental abundances, and a covariance matrix that combines experimental and theoretical uncertainties. The measured values and uncertainties of Y$_{\textrm{P}}$~\cite{Zyla:2020zbs} and D/H~\cite{Cooke:2017cwo} are
\begin{eqnarray}
\label{eq:BBN2}
\textrm{Y}_{\textrm{P}} &=& 0.245 \pm 0.003 \,, \\ \nonumber
\text{D/H} &=& (2.527 \pm 0.030) \times 10^{-5} \, .
\end{eqnarray}
Theoretical uncertainties and correlations between predicted Y$_\text{P}$ and D/H values arise from uncertainties in nuclear rates. The D/H theoretical uncertainty varies with both $\boldsymbol \theta$ and $\Omega_b h^2$, and can be comparable to or even exceed the measurement uncertainty; it is therefore evaluated at each parameter point. We compute the theoretical uncertainty and correlations of both D/H and Y$_\text{P}$ by varying the nuclear rates by 1$\sigma$, and adding the resulting fractional variations to the abundances in quadrature. Finally, both theoretical and measurement uncertainties are added in quadrature to obtain the full BBN covariance matrix. Further details on this procedure are provided in Appendix B\@. 

We then calculate the profile likelihood ratio $\lambda_p(\boldsymbol{\theta}) = L_p(\boldsymbol{\theta}) / \hat{L}_p$, where $L_p(\boldsymbol \theta) = \max_{\Omega_b h^2} L(\boldsymbol \theta, \Omega_b h^2)$ and $\hat{L}_p = \max_{\boldsymbol{\theta}, \Omega_b h^2} L(\boldsymbol \theta, \Omega_b h^2)$. By Wilk's theorem, the quantity $-2 \log \lambda_p(\boldsymbol{\theta})$ follows a chi-squared distribution with degrees of freedom given by the number of model parameters~\cite{Zyla:2020zbs}. 68~(95)\% confidence limits for $\boldsymbol \theta$ in our two-parameter benchmark model are therefore set when  $-2 \log \lambda_p(\boldsymbol{\theta}) = 2.30~(6.18)$.

\vspace{0.1in}
\noindent \textbf{Dark Photon Model Constraints.---}Fig.~\ref{fig:Joint_Constraint_Blobs} illustrates the interplay between CMB and BBN measurements in constraining a complex scalar $\chi$ and a dark photon with  $m_{A'} = 3 m_\chi$.  The purple~(pink) lines correspond to the constraints on $m_\chi$ and $\Delta N_\nu$ that are consistent with the Planck (projected Simons Observatory) measurements.  Above $m_\chi \sim 20$~MeV, the predicted value of \Neff\, approaches the standard cosmological value because the DM freezes out well before neutrino decoupling, and so the DM annihilations heat the electromagnetic and neutrino sectors equally.

The impact of the dark sector becomes apparent when $m_{\chi}\lesssim \SI{20}{\mega\eV}$, and entropy is injected into the electromagnetic sector during and after the period of neutrino decoupling, when the SM bath has a temperature of $T_{\nu d} \sim \SI{2}{\mega\eV}$.  In this case, \Neff\, decreases relative to the standard value because the electromagnetic sector is preferentially heated, but the photon temperature today is fixed at its measured value of 2.7 K\@.  A nonzero $\Delta N_\nu$ can restore \Neff\, to its measured value; when  $m_\chi$ falls below $\sim \SI{20}{\mega\eV}$, the fit clearly prefers a larger $\Delta N_\nu$ to explain the observed value of \Neff. Below $m_\chi \lesssim \SI{1}{\mega\eV}$, the DM is relativistic throughout neutrino decoupling, entropy is dumped only into the electromagnetic sector regardless of mass, and the constraints level off (Fig.~\ref{fig:Joint_Constraint_Blobs} inset).

 The result of including BBN constraints from Y$_\text{P}$ and D/H is indicated by the shaded regions in Fig.~\ref{fig:Joint_Constraint_Blobs}.  BBN clearly adds significant discriminating power, placing a 95\% lower confidence bound on the DM mass of $m_\chi \sim \SI{5}{\mega\eV}$ when combined with Planck data, regardless of $\Delta N_\nu$. The Simons Observatory will have improved sensitivity to $m_{\chi}$ with its more precise measurement of $N_\text{eff}$, while also reducing the uncertainty on $\Delta N_{\nu}$. 

The introduction of an \SI{}{\mega\eV}-scale DM particle and dark photon leads to a variety of effects on BBN physics, which are summarized in Fig.~\ref{fig:BBN_only} for fixed $\Omega_b h^2 = 0.022358$.  The solid (dashed) colored lines correspond to different ratios of $m_{A'}/m_\chi$ for $\Delta N_{\nu} = $ 0 (1).  The case where $m_{A'}/m_\chi \rightarrow \infty$ is consistent with Refs.~\cite{Kolb_1986, Serpico:2004nm, Boehm:2013jpa,Nollett:2013pwa}.  The interplay of the following four quantities is relevant for understanding this behavior: \textit{i)}~the neutron-proton ratio, which is positively correlated with the helium-4 abundance, \textit{ii)}~the baryon-to-photon ratio, $\eta$, which is inversely correlated with the deuterium abundance,  \textit{iii)}~the expansion rate, which impacts both the neutron abundance and the deuterium burning rate~\cite{Mukhanov:2003xs}, and \textit{iv)}~the rate of neutron-proton interconversion, affected by a modified $T_\nu$ (at fixed $T_\gamma$).  When $m_\chi \gtrsim 3\,T_{\nu d}$, BBN proceeds as per the standard scenario.  When $m_e \lesssim m_\chi \lesssim 3\,T_{\nu d}$, DM injects significant entropy into the electromagnetic sector after neutrino decoupling.  The expansion rate is therefore slower at fixed photon temperature, which drives down the deuterium abundance as there is more time to convert deuterium to heavier elements.  Meanwhile, near-cancellation of effects on neutron-proton interconversion and on the expansion rate keeps Y$_{\textrm{P}}$ essentially constant in this regime~\cite{Nollett:2013pwa}.  When $m_\chi \lesssim m_e$, the DM acts as a new relativistic species during BBN\@.  This increases the expansion rate, causing weak interactions to decouple earlier, thereby increasing Y$_{\textrm{P}}$.  In contrast, D/H is further reduced because post-BBN DM annihilations lead to an increased $\eta$ during BBN\@.

$\Delta N_\nu > 0$ compounds the effect of introducing an MeV-scale DM particle by further increasing the expansion rate.  This increases the production of helium-4 and mitigates the decrease in the deuterium abundance.  As a result, an increase in $\Delta N_\nu$ shifts the $\Delta N_\nu = 0 $ curves in Fig.~\ref{fig:BBN_only} upwards.   

As $m_{A'}/m_\chi$ is reduced, the effects described above are only further enhanced because of the additional entropy injection from the dark photon.  In particular circumstances, the presence of the dark photon can qualitatively affect the shape of the D/H and Y$_{\textrm{P}}$ curves in Fig.~\ref{fig:BBN_only}.  For example, the $m_{A'}/m_\chi =10$ curve exhibits distinctive behavior when $m_\chi \lesssim m_e$.  The observed plateau in D/H corresponds to the transition from the point where the dark photon entropy injection heats the photon bath, to the point where the dark photon acts as an additional relativistic species throughout BBN\@.  Elsewhere, the curves for different values of $m_{A'}/m_\chi$ look similar, but the curves shift to the right as $m_{A'}/m_\chi$ decreases, since the entropy injection from the dark photon increases as $m_{A'}$ decreases.

The current measurements of D/H and Y$_{\textrm{P}}$ are indicated in Fig.~\ref{fig:BBN_only}.  For the case where $m_\chi \rightarrow \infty$, we find a $\sim 2\sigma$ discrepancy with the measured deuterium abundance from Ref.~\cite{Cooke:2017cwo}, consistent with other studies using \texttt{PRIMAT}.  Refs.~\cite{Yeh:2020mgl, Pisanti:2020efz}, which perform independent analyses with different code packages, find better agreement with larger uncertainties.  These differences are likely due, at least in part, to differing treatments of the 2D~$\rightarrow$~$^3$He~+~n  and 2D $\rightarrow$ $^3$H + p reactions~\cite{Pisanti:2020efz,Pitrou:2020etk}.  The method presented in this Letter can be easily adapted to account for future improvements in BBN calculations.

Table~\ref{tab:mratio_results}~enumerates the 95\% confidence lower bound on $m_\chi$ for a complex scalar, a Majorana fermion, and a Dirac fermion for different values of $m_{A'}/m_\chi$.  In all cases, the minimum mass is constant for $m_{A'}/m_\chi \gtrsim 10$ and is a factor of 1.3--1.8 weaker than joint CMB and BBN limits assuming $\Delta N_\nu = 0$~\cite{sabti2021implications}. We find a robust lower bound of $m_\chi > \SI{3.9}{\mega\eV}$ across all DM particle types, for any non-zero $\Delta N_{\nu}$, using Planck data~\cite{Aghanim:2018eyx}.  The Simons Observatory will be sensitive to heavier masses by several \SI{}{\mega\eV}.

The lower bound on $m_\chi$ has important implications for experiments searching for dark sector DM\@. Our benchmark model is commonly used to present bounds and sensitivity projections for direct-detection and accelerator-based experiments.  To date, the generality of CMB $N_\text{eff}$ limits on this model has been questioned due to the degeneracy between DM entropy injection and $\Delta N_\nu$~\cite{Boehm:2013jpa,Essig:2015cda,Wilkinson:2016gsy,Izaguirre:2017bqb,Berlin:2018bsc}.  Because our joint CMB and BBN constraints apply for any $\Delta N_\nu > 0$, they address these prior concerns and establish a robust cosmological bound on the dark sector model under consideration.

\begin{table}[t]
\footnotesize
\begin{center}
\begin{tabular}{C{1.8cm} | C{1.9cm}C{1.9cm}C{1.9cm}}
  \Xhline{3\arrayrulewidth}
\renewcommand{\arraystretch}{1}
   &  \multicolumn{3}{c}{\textbf{Minimum $\boldsymbol{m_\chi}$ [MeV]}} \\
$m_{A'}/m_{\chi}$  & Complex Scalar     &  Majorana Fermion & \,\,Dirac \,\, Fermion   \Tstrut\Bstrut   \\   
 \hline
 0.5 & 14.3\,\,(16.5) & 14.3\,\,(16.5) & 15.2\,\,(17.1) \\
 1  & 9.0\,\,(10.1)  & 9.0\,\,(10.1)  & 10.4\,\,(11.5) \\
 1.5  & 7.1\,\,(8.1)  & 7.1\,\,(8.0)  & 9.0\,\,(10.0) \\
 3  & 5.2\,\,(6.2)  & 5.0\,\,(6.1)  & 7.9\,\,(9.1)  \\
 $\geq 10$ & 4.3\,\,(5.8)  & 4.0\,\,(5.6)  & 7.8\,\,(9.1)  \\
  \Xhline{3\arrayrulewidth}
\end{tabular}
\end{center}
\caption{The joint Planck CMB and BBN 95\% lower limit on the mass of a DM particle that is a complex scalar, Majorana fermion, or Dirac fermion in our benchmark model.  The mass limits are provided for different ratios of the dark photon to DM mass.  The values in parentheses are the projected Simons Observatory and BBN constraints~\cite{SimonsObservatory:2018koc,Sabti:2019mhn}.  The value of $\Omega_b h^2$ that maximizes the likelihood was chosen for each parameter point.} 
\label{tab:mratio_results}
\end{table}

\begin{figure}[t!]
    \centering
    \includegraphics[width=0.49\textwidth]{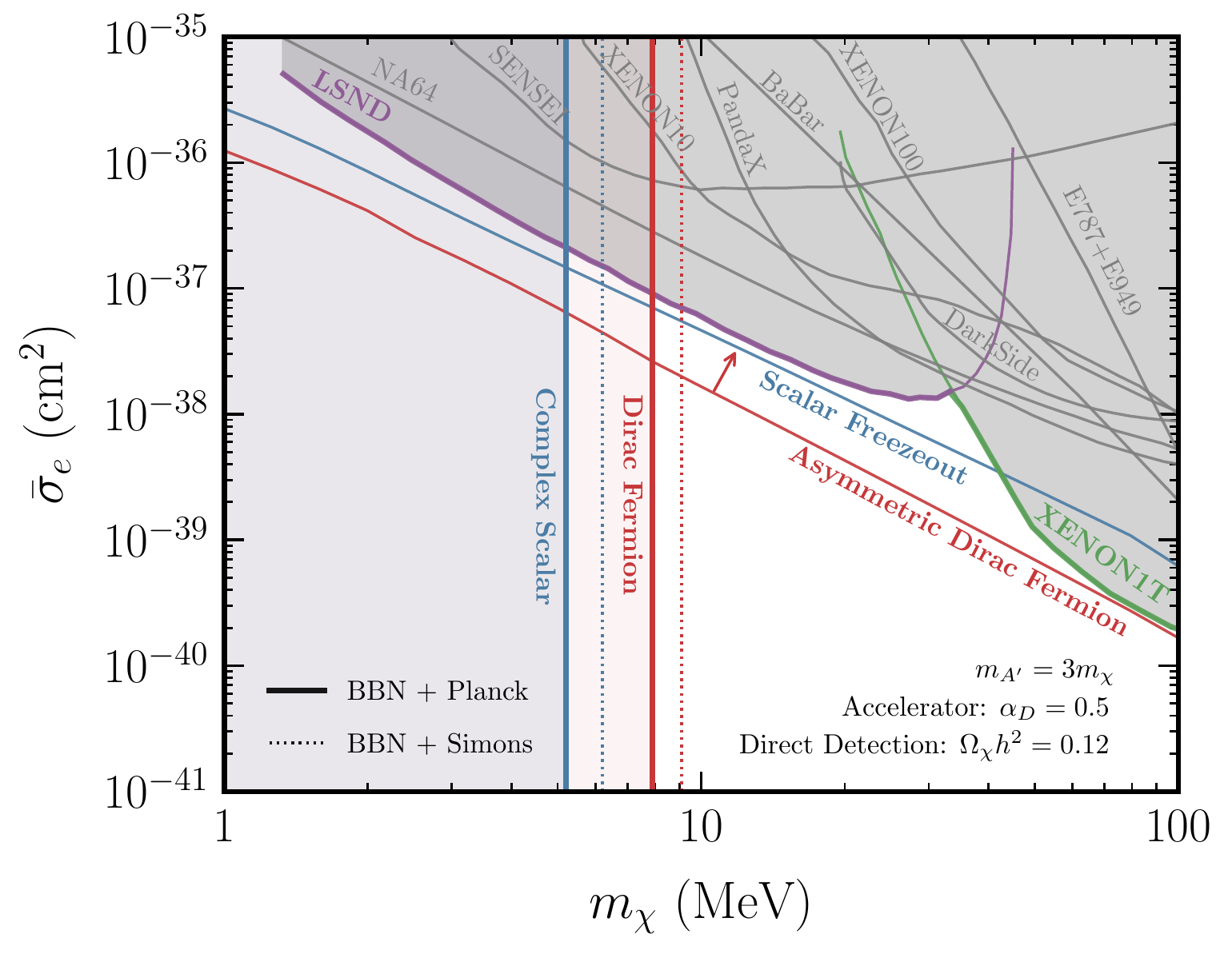}
    \caption{Joint CMB and BBN 95\% confidence constraints on $m_\chi$ when $\chi$ is a Dirac fermion (shaded red) or a complex scalar (shaded blue), with $m_{A'} = 3 m_\chi$. The $y$-axis shows the reference dark matter-electron scattering cross section $\overline{\sigma}_e$. Projected constraints from the Simons Observatory are demarcated by the vertical dotted lines.  The blue line indicates where a complex scalar $\chi$ undergoes a standard freezeout to attain the correct relic abundance; the line above which the freezeout of an asymmetric Dirac fermion $\chi$ can produce all of DM is indicated in red. The region of parameter space that is ruled out by experiments at the 90\% confidence level is shaded grey, with the strongest constraints coming from LSND~\cite{deNiverville:2011it} (purple) and XENON1T~\cite{Aprile:2019xxb} (green). Other experimental limits are shown as labeled gray lines~\cite{Adler:2002hy,Artamonov:2009sz,Essig:2012yx,Banerjee_2019,Lees:2017lec,Essig:2017kqs,Agnes:2018oej,Barak:2020fql,Cheng:2021fqb}. All direct-detection limits on $\overline{\sigma}_e$ assume $\chi$ is all of the DM, while accelerator-based limits are independent of the $\chi$-abundance, but assume $\alpha_D = 0.5$. }
    \label{fig:Exp_plot}
\end{figure}

We present our results in Fig.~\ref{fig:Exp_plot} for $m_{A'} = 3 m_\chi$ in terms of the reference DM-electron scattering cross section, $\overline{\sigma}_e \equiv 16 \pi \alpha \alpha_D \epsilon^2 \mu_{\chi e}^2 / (\alpha^2 m_e^2 + m_{A'}^2)^2$, where $\alpha$, $\alpha_D$ are the electromagnetic and dark sector fine structure constants respectively, $\epsilon$ is the SM-$A'$ mixing parameter, and $\mu_{\chi e}$ is the electron-$\chi$ reduced mass.  We show the lower limit on $m_\chi$ for a Dirac fermion and a complex scalar, together with existing direct-detection~\cite{Essig:2012yx,Essig:2017kqs,Agnes:2018oej,Aprile:2019xxb,Barak:2020fql,Cheng:2021fqb} and accelerator~\cite{Adler:2002hy,Artamonov:2009sz,deNiverville:2011it,Batell:2014mga,Banerjee_2019,Lees:2017lec} limits on $\overline{\sigma}_e$, assuming $\chi$ makes up all of the DM for direct-detection experiments, and choosing $\alpha_D = 0.5$ for beam experiments. We solve the Boltzmann equation for $\chi$ with the processes $\chi \overline{\chi} \leftrightarrow e^+e^-$ and $\mu^+\mu^-$ to obtain \textit{i)} $\overline{\sigma}_e$ as a function of $m_\chi$ for a symmetric complex scalar $\chi$ undergoing a standard freezeout through annihilation into SM fermions, and \textit{ii)} the lower limit on $\overline{\sigma}_e$ as a function of $m_\chi$ for an asymmetric Dirac fermion $\chi$ freezing out through annihilation into SM fermions, given the Planck limits on DM annihilation~\cite{Lin:2011gj,Slatyer:2015jla,Aghanim:2018eyx} (see, e.g., Refs.~\cite{Steigman:2012nb,Feng:2017drg,Berlin:2018bsc} for similar relic abundance calculations). The joint constraints set a lower limit of $m_\chi > \SI{5.2}{\mega\eV}$ (\SI{7.9}{\mega\eV}) for a complex scalar (Dirac fermion) $\chi$ with an arbitrary number of light degrees of freedom at 95\% confidence. The Simons Observatory~\cite{SimonsObservatory:2018koc,Sabti:2019mhn} is forecasted to have slightly improved sensitivity.

Our limits apply only for dark sectors in chemical equilibrium with the SM while $A'$ is relativistic, where our entropy injection calculation is valid. In the early Universe, processes such as $\chi \overline{\chi} \leftrightarrow e^+e^-$ bring the dark sector into chemical equilibrium with the SM while $T_\gamma \gtrsim m_{A'}$ for sufficiently large $\overline{\sigma}_e$.  We estimate this $\overline{\sigma}_e$ by requiring the rate of $e^+e^- \to \chi \overline{\chi}$ to exceed the Hubble rate at $T_\gamma = m_{A'}$, and find $\overline{\sigma}_e \gtrsim \SI{e-46}{\centi\meter\squared} (\SI{10}{\mega\eV}/m_{\chi})^3$. 

\vspace{0.1in}
\noindent
\textbf{Conclusions.---} We have developed a method for obtaining joint CMB and BBN constraints on general dark sectors.  As a concrete example, we focused on the dark sector model where a DM particle interacts with the SM through a massive dark photon mediator, including the possibility of an arbitrary number of light degrees of freedom. We place a 95\% confidence  lower bound of $m_\chi \gtrsim \SI{4}{\mega\eV}$ on the DM mass as long as the dark sector is fully thermalized with the SM in the early Universe.  In Table~\ref{tab:mratio_results}, we also illustrate how the constraints strengthen with decreasing dark photon mass.  Recent studies have identified cosmological and astrophysical probes of $\chi e\rightarrow\chi e$ scattering~\cite{Buen-Abad:2021mvc,Nguyen:2021cnb}, resulting in constraints on $\bar{\sigma}_e$ that are many orders of magnitude weaker than the range plotted in Fig.~\ref{fig:Exp_plot} (e.g., $\overline{\sigma}_e \gtrsim \SI{e-30}{\centi\meter\squared}$ for $m_\chi = \SI{1}{\mega\eV}$). Thus, for the example of $m_\chi = \SI{1}{\mega\eV}$, our constraints are expected to apply between $\SI{e-43}{\centi\meter\squared} \lesssim \overline{\sigma}_e \lesssim \SI{e-30}{\centi\meter\squared}$, above which $\chi e \to \chi e$ becomes important. To our knowledge, there are no existing models of electrophilic DM that completely evade our bounds, though model-building extensions have been proposed for other scenarios that can weaken the CMB $N_\text{eff}$ constraints and allow for \SIrange{1}{10}{\mega\eV} electromagnetically-coupled dark sector particles, e.g., by allowing some DM annihilation into neutrinos~\cite{Escudero_2019,Sabti:2019mhn,Agashe:2020luo}.  We hope to better understand the robustness of these cosmological limits on generic dark sectors in future work. 

Appendices~\ref{app:modifications_to_codes} and~\ref{app:integrals} describe the modifications made to \texttt{nudec\_BSM} and \texttt{PRIMAT} to handle the dark sector model studied in this Letter.  The modified code is available upon request from the authors.

\vspace{0.1in}
\noindent
{\bf Acknowledgements.---}The authors thank Kaustubh Agashe, Alexandre Arbey, Asher Berlin, Kimberly Boddy, Manuel Buen-Abad, Bhaskar Dutta, Rouven Essig, Jonathan Feng, Vera Gluscevic, David McKeen, Siddharth Mishra-Sharma, Cyril Pitrou, Maxim Pospelov, Jordan Smolinsky, Yuhsin Tsai, Neal Weiner, Tien-Tien Yu and Yiming Zhong for fruitful conversations.  This material is based upon work supported by the NSF Graduate Research Fellowship under Grant No. DGE1839302.  HL and ML are supported by the DOE under Award Number DE-SC0007968.  ML is also supported by the Cottrell Scholar Program through the Research Corporation for Science Advancement. HL is also supported by NSF grant PHY-1915409, and the Simons Foundation.  JTR is supported by NSF CAREER grant PHY-1554858 and NSF grant PHY-1915409.  This work was performed in part at the Aspen Center for Physics, which is supported by NSF grant PHY-1607611.  The work presented in this paper was performed on computational resources managed and supported by Princeton Research Computing. This research made extensive use of the publicly available codes \texttt{nudec\_BSM}~\cite{Escudero_2019,Escudero:2020dfa} and \texttt{PRIMAT}~\cite{Pitrou:2018cgg,Pitrou:2020etk}, as well as the  
\texttt{IPython}~\citep{PER-GRA:2007}, 
\texttt{Jupyter}~\citep{Kluyver2016JupyterN}, \texttt{matplotlib}~\citep{Hunter:2007}, 
\texttt{NumPy}~\citep{numpy:2011}, and
\texttt{SciPy}~\citep{Jones:2001ab} software packages.

\bibliography{references}

\clearpage

\onecolumngrid
\appendix
\makeatletter

\label{supp}

\noindent

\section{\Neff\,Constraint}\label{app:NeffDerivation}

\setcounter{equation}{0}
\setcounter{figure}{0}
\setcounter{table}{0}
\renewcommand{\theequation}{A\arabic{equation}}
\renewcommand{\thefigure}{A\arabic{figure}}
\renewcommand{\thetable}{A\arabic{table}}

This appendix provides a pedagogical derivation of \Neff\, as a function of $m_{\chi}$ in the presence of a dark matter~(DM) particle $\chi$ and a dark photon $A'$.  We assume that neutrino decoupling is instantaneous in this appendix so that the results are analytical; however, the results in the main Letter include corrections for non-instantaneous neutrino decoupling, obtained using the modified \texttt{nudec\_BSM} code described in Appendices~\ref{app:modifications_to_codes} and~\ref{app:integrals}.  

Our goal is to use conservation of entropy to determine the increase in photon temperature due to DM freezeout.  The entropy density of a given species $i$ is defined as
\begin{equation}
s_i=\frac{\rho_i+P_i}{T_i} \,, \nonumber\label{eq:entropy_density}
\end{equation}
where $\rho_i$ is the energy density, $P_i$ is the pressure, and $T_i$ is the temperature of species $i$.\footnote
{
We assume that the chemical potentials of all species are zero, and that the energy density of each neutrino species is identically given by $\rho_{\nu}$.
}
For the benchmark dark sector considered in this work, the total radiation density of the early Universe is given by
\begin{equation}
    \rho_R=\rho_{\gamma}+\rho_e+3\rho_{\nu}+\rho_{\xi}+\rho_{\chi}+\rho_{A'} \,,\label{eq:rad_density_WIMP}\nonumber
\end{equation}
where we have included the energy density of additional light degrees of freedom $\xi$, in addition to the contribution from photons, electrons/positrons, neutrinos, $\chi$, and $A'$.
The energy density of species $i$ with energy $E_i(p)$ is given by
\begin{equation}
\rho_i(E_i) =\frac{g_i}{2\pi^2}\int_0^\infty \textrm{d}p\,f_i(p)\,E_i(p)\,p^2 \,,\nonumber\label{eq:energy_density_general}
\end{equation}
where the distribution function $f_i(p)$ is either Bose-Einstein or Fermi-Dirac, and $g_i$ corresponds to the internal degrees of freedom of the particle (i.e., 1 for a real scalar, 2 for a complex scalar or Majorana fermion, and 4 for a Dirac fermion).  Taking $E_i(p)=\sqrt{p^2+m_i^2}$ and substituting in for $f_i(p)$ yields
\begin{equation}
\rho_i(m_i, T_i) =\frac{g_i}{2\pi^2}\int_0^\infty\,\textrm{d}p\,\frac{p^2\sqrt{p^2+m_i^2}}{e^{\sqrt{p^2+m_i^2}/T_i}\pm 1} \, .\nonumber
\end{equation}
For notational convenience, we change variables to $x_i\equiv m_i/T_i$ and $\lambda_i \equiv p/T_i$, so that the energy density becomes
\begin{equation}
\rho_i(x_i)=\frac{g_i}{2\pi^2}\, T_i^4 \, J_\pm(x_i) \,\label{eq:rho_x} \, ,
\end{equation}
where
\begin{equation}
J_\pm(x)\equiv\int_0^\infty\dd\lambda\, \frac{\lambda^2\sqrt{\lambda^2+x^2}}{e^{\sqrt{\lambda^2+x^2}}\pm1} \, .\nonumber 
\nonumber\label{eq:defJ}
\end{equation}
Meanwhile, the pressure of species $i$ is given by
\begin{equation}
P_i(E_i) =\frac{g_i}{(2\pi)^2}\int_0^{\pi}\dd\theta \sin\theta \int_0^\infty \textrm{d}p\, f_i(p)\, \frac{p^2}{3E_i(p)}\,p^2\,,\nonumber
\end{equation}
which can be rewritten as
\begin{equation}
P_i(x_i)=\frac{g_i}{2\pi^2}\, T_i^4\int_0^\infty\dd\lambda\, \frac{\lambda^4}{3\sqrt{\lambda^2+x_i^2}}\frac{1}{e^{\sqrt{\lambda^2+x_i^2}}\pm 1} \,.
\label{eq:pressure_x}
\end{equation}
The total entropy density is therefore
\begin{equation}
s_{\textrm{tot}} =\sum_{i}\frac{g_i}{2\pi^2} \, T^3_i\left[ J_{\pm}(x_i)+\int_0^\infty\dd \lambda\,\frac{\lambda^4}{3\sqrt{\lambda^2+x_i^2}}\frac{1}{e^{\sqrt{\lambda^2+x_i^2}}\pm 1}\right] \,.
\label{eq:entropy_full}
\end{equation}

For light or massless species ($x_i \ll 1$), \Eq{eq:entropy_full} can be further simplified because $J_\pm(x)$ can be evaluated analytically in the relativistic limit.  In this case,
\begin{align}
J_+(0)=\frac{7\pi^4}{120} \quad \text{ and } \quad J_-(0)=\frac{\pi^4}{15}\nonumber\,,
\end{align}
which gives
\begin{equation}\label{eq:energy_density_boson_fermion}
  \rho_i(0)=\frac{\pi^2}{30} \, g_i \, T_i^4 \times
  \begin{cases}
                                   1 & \text{for bosons} \\
                                   \frac{7}{8} & \text{for fermions} \, . \\
  \end{cases}
\end{equation}
We define
\begin{equation}
g_*(T_{\gamma})\equiv\sum_{i\in\textrm{bosons}}g_i\left(\frac{T_i}{T_{\gamma}}\right)^4+\frac{7}{8}\sum_{i\in\textrm{fermions}}g_i\left(\frac{T_i}{T_{\gamma}}\right)^4\nonumber\label{eq:gstar},
\end{equation}
where $T_{\gamma}$ is the photon temperature, such that the total energy density of all relativistic SM species and DM is
\begin{equation}
\rho_{\textrm{tot,rel}}=\frac{\pi^2}{30}\, T_{\gamma}^4 \, g_*(T_{\gamma}) \, .\nonumber\label{eq:rho_0}
\end{equation}
The expression for pressure also simplifies in this limit; taking $x\rightarrow0$ in the relativistic limit, we find
\begin{equation}
P_i(0)=\frac{g_i}{6\pi^2} \, T_i^4 \int_0^\infty \textrm{d}\lambda\, \frac{\lambda^3}{e^\lambda\pm1} =\frac{1}{3}\, \rho_i(0) \, ,\nonumber
\end{equation}
returning the expected relation for a relativistic gas.  The total entropy density for relativistic species then becomes
\begin{align}
s_{\textrm{tot,rel}} =\frac{4}{3}\sum_i\frac{\rho_i(0)}{T_i}=\frac{2\pi^2}{45} \, T_{\gamma}^3\left[\sum_{i\in\textrm{bosons}} g_i\left(\frac{T_i}{T_{\gamma}}\right)^3+\frac{7}{8}\sum_{i\in\textrm{fermions}} g_i\left(\frac{T_i}{T_{\gamma}}\right)^3\right] \, .\label{eq:entropy_sum1}
\end{align}
We can simplify this expression by defining
\begin{equation}
g_{*S}(T_{\gamma})\equiv\sum_{i\in\textrm{bosons}}g_i\left(\frac{T_i}{T_{\gamma}}\right)^3+\frac{7}{8}\sum_{i\in\textrm{fermions}}g_i\left(\frac{T_i}{T_{\gamma}}\right)^3\nonumber\label{eq:gstarS}
\end{equation}
such that Eq.~\eqref{eq:entropy_sum1} becomes
\begin{equation}
s_{\textrm{tot,rel}}=\frac{2\pi^2}{45}\, T_{\gamma}^3 \, g_{*S}(T_{\gamma})\label{eq:entropy_sum} \, .
\end{equation}
To find the total entropy density including contributions from nonrelativistic species, we must evaluate the integrals in Eq.\eqref{eq:entropy_full} numerically.

Now that we have an expression for entropy density as a function of temperature and particle mass, we can use conservation of entropy to determine the photon heating from DM freeze out.  We assume that electrons, photons, and DM are in thermal equilibrium until they decouple from the SM plasma.  At the point of neutrino decoupling, the total entropy density of all radiative species
\begin{equation}
    s_{\textrm{tot}}\big|_{a_d}=\frac{2\pi^2}{45} \, T_{\nu d}^3\left[g_\gamma+\frac{7}{8}\big(g_e \phi(x_{ed})+3g_\nu\big)+\tilde{g}_{\chi}\phi(x_{\chi d})+g_{A'}\phi(x_{A'd})\right] + \frac{2\pi^2}{45}T^3_{\xi,a_d}\tilde{g}_{\xi}\, ,
    \label{eq:total_entropy_a1}
\end{equation}
where $T_{\nu d}$ is the common temperature of all SM species and dark sector species at the point of neutrino decoupling,\footnote{Typically, we may assume $T_{\nu d}\approx\SI{2}{MeV}$ for these analytical calculations, though this assumption is not enforced to calculate the constraints in the main Letter.  \texttt{nudec\_BSM} allows us to eliminate the assumption of instantaneous neutrino decoupling, and the neutrino decoupling temperature does not need to be known \textit{a priori} to obtain these constraints.} which occurs at the corresponding scale factor $a_d$, and 
\begin{equation}
  \phi(x_{id})\equiv\frac{s_i(x_{id})}{s_i(0)} \,,\nonumber
\end{equation}  
where $x_{id} \equiv m_i / T_{\nu d}$ for species $i$. Following Ref.~\cite{Nollett_2014b}, we have introduced the notation $\tilde{g}_{\chi}$ to absorb fermion factors of 7/8 for particles with different numbers of degrees of freedom, i.e., $\tilde{g}_{\chi}=1,\, 7/4,\, 2$, and $7/2$ for a real scalar, Majorana fermion, complex scalar, and Dirac fermion, respectively.\footnote{Since we restrict to the case of a massive vector boson, there is no need to define a corresponding $\tilde{g}_{A'}$.  However, additional variables $\tilde{g}_i$ may be required to accommodate more complex dark sectors.}  We also include a term $\tilde{g}_{\xi}$ to parametrize the entropy contributed by equivalent neutrinos (which also may have additional fermion factors attached), as well as the temperature of the equivalent neutrinos at the point of neutrino decoupling $T_{\xi,a_d}$, which is not necessarily the temperature of the SM plasma.

\par
After some time, the DM and the positrons/electrons annihilate, injecting energy into the photon bath.  DM annihilations that occur after neutrino decoupling inject entropy into the electromagnetic plasma only, thereby increasing  $T_{\gamma}$; the same occurs during electron-positron annihilation.  In this Appendix, we assume that $e^+e^-$ annihilation occurs entirely after neutrinos have decoupled, so that all of the annihilations heat the electromagnetic sector and our calculations remain analytical.  In reality, there is some overlap between neutrino decoupling and electron-positron annihilation that causes neutrinos to be slightly heated by this process as well; this effect is tracked numerically in the main Letter.  To track temperature evolution during these processes, we separately conserve entropy among all species coupled to photons (photons, electrons, DM and dark photons), and among all species that do not receive an energy injection from electron-positron annihilation or DM freeze out.  This means that, at some reference value of the scale factor $a=a_2$ after electron-positron annihilation and DM freeze out, the total entropy is given by 
\begin{equation}
    s_{\textrm{tot}}\big|_{a_2}=\frac{2\pi^2}{45}\left[g_\gamma T_{\gamma, a_2}^3+\frac{21}{8}g_\nu T_{\nu, a_2}^3+g_{\xi}T^3_{\xi,a_2}\right] \, ,
    \label{eq:total_entropy_a2}
\end{equation}
where $T_{\gamma, a_2}\neq T_{\nu, a_2}$ because photons are heated after neutrino decoupling.  Because neutrinos decouple completely before any other species freezes out in this approximation, we expect $T_{\nu}$ to scale as $a^{-1}$, i.e., $a_dT_{\nu d} =a_2 T_{\nu, a_2}$.  A similar relation holds for the equivalent neutrino temperature.  Equating the total entropy before and after electron-positron annihilation and DM freeze out then yields
\begin{equation}
    g_\gamma+\frac{7}{2}\phi(x_{ed})+\frac{21}{8}g_\nu+\tilde{g}_{\chi}\phi(x_{\chi d})+g_{A'}\phi(x_{A'd})+g_{\xi}=\left(\frac{T_\gamma}{T_\nu}\right)_0^3g_\gamma+\frac{21}{8}g_\nu+g_{\xi} \, ,\nonumber
\end{equation}
or
\begin{equation}
    \left(\frac{T_{\nu}}{T_{\gamma}}\right)^3_0=\frac{g_{\gamma}}{g_{\gamma}+\frac{7}{2}\phi(x_{ed})+\tilde{g}_{\chi}\phi(x_{\chi d})+g_{A'}\phi(x_{A' d})} \, . \label{eq:temperature_ratio}
\end{equation}
We use the subscript `0' to denote a late point in time when $T_\gamma \ll m_\chi$, $m_{A'}$, and $m_e$.  Knowing the late-time ratio of the neutrino and photon temperatures, we can also use Eq.~\eqref{eq:energy_density_boson_fermion} to find the ratio of their radiation densities.  This allows us to find the effective number of relativistic degrees of freedom,
\begin{equation}
    N_{\textrm{eff}}\equiv \left(\frac{\rho_R-\rho_{\gamma}}{\rho_{\nu,\textrm{std}}}\right)_0 \,.
\end{equation}
Here, the subscript ``std" denotes the standard result (i.e., ignoring contributions from DM freeze out or other dark sector effects), including the assumption of \emph{i)} instantaneous neutrino decoupling, and \emph{ii)} no heating of the neutrinos from electrons and positrons.  Using Eq.~\eqref{eq:energy_density_boson_fermion}, we may write this as
\begin{equation}
    \left(\rho_{\nu,\textrm{std}}\right)_0 = \frac{7}{8}\left(\frac{T_{\nu}}{T_{\gamma}}\right)_{0,\textrm{std}}^4\left(\rho_{\gamma}\right)_0=\frac{7}{8}\left(\frac{4}{11}\right)^{4/3}\left(\rho_{\gamma}\right)_0\nonumber,
\end{equation}
where we have used expressions similar to Eqs.~\eqref{eq:total_entropy_a1} and~\eqref{eq:total_entropy_a2} to find the ratio of the photon and neutrino temperatures under the assumptions listed above (see~\cite{Dodelson:2003ft}).
We can also write
\begin{align}
\left(\frac{\rho_R}{\rho_{\gamma}}\right)_0&=1+\left(\frac{3\rho_{\nu}}{\rho_\gamma}\right)_0+\left(\frac{\rho_\xi}{\rho_\gamma}\right)_0\nonumber\\
&=1+\left(\frac{\rho_\nu}{\rho_\gamma}\right)_0\big(3+\Delta N_\nu\big),
\end{align}
with $\Delta N_\nu\equiv\left(\frac{\rho_\xi}{\rho_\nu}\right)_0$ as the number of equivalent neutrinos in the early Universe.  With the result from Eq.~\eqref{eq:temperature_ratio}, \Neff\ is given by
\begin{equation}
    N_{\textrm{eff}}=3\left(\frac{11}{4}\frac{g_{\gamma}}{g_{\gamma}+\frac{7}{2}\phi(x_{ed})+\tilde{g}_{\chi}\phi(x_{\chi d})+g_{A'}\phi(x_{A' d})}\right)^{4/3}\left(1+\frac{\Delta N_\nu}{3}\right).
\end{equation}
This provides an analytical expression to find \Neff\, in the presence of DM with mass $m_{\chi}$ and a dark photon with mass $m_{A'}$.  This result can be easily extended to accommodate richer dark sectors.  In order to account for additional corrections from non-instantaneous neutrino decoupling and from QED, numerical tools are required; the modifications made to \texttt{nudec\_BSM} to accomplish this are detailed in Appendix~\ref{app:modifications_to_codes}\@.  This code eliminates any need for an assumption of instantaneous neutrino decoupling by explicitly tracking the photon and neutrino temperatures as a function of time with a pair of coupled Boltzmann equations, taking into account collisions that occur between electrons and neutrinos and calculating the resulting energy transfer between the two sectors as neutrinos decouple over a finite period of time~\cite{Escudero_2019}.

\noindent

\section{Treatment of CMB and BBN Covariances}\label{app:joint_constraint_details}

\setcounter{equation}{0}
\setcounter{figure}{0}
\setcounter{table}{0}
\renewcommand{\theequation}{B\arabic{equation}}
\renewcommand{\thefigure}{B\arabic{figure}}
\renewcommand{\thetable}{B\arabic{table}}

In this appendix, we detail how to obtain the joint CMB and BBN constraints outlined in the main body. Starting from any dark sector model with some model parameters $\boldsymbol\theta$, we compute the overall effect of the dark sector on \textit{i)} $N_\text{eff}$, the effective number of relativistic degrees of freedom at late times, and \textit{ii)} the primordial abundances of elements after the process of BBN has ended. For each dark sector model, we perform the abundance calculations over the range $\Omega_b h^2 \in [0.0218,0.0226]$, which is much broader than the Planck uncertainty on this parameter~\cite{Aghanim:2018eyx}.  We leave a discussion of how to implement this procedure using existing numerical codes to the following section, simply assuming for now that the parameters $N_\text{eff}(\boldsymbol\theta)$, $\textrm{Y}_{\textrm{P}}(\boldsymbol\theta, \Omega_b h^2)$, and D/H$(\boldsymbol\theta, \Omega_b h^2)$ can be obtained, together with theoretical uncertainties and correlations affecting the BBN calculation. 

To assess the consistency of a given dark sector model with experimental measurements of primordial elemental abundance and the CMB, we perform a hypothesis test on our model parameters by constructing a profile likelihood ratio, as discussed in the main text. We define
\begin{alignat}{1}
    L(\boldsymbol\theta, \Omega_b h^2) = L_\text{BBN}(\boldsymbol\theta, \Omega_b h^2) \cdot L_\text{CMB}(\boldsymbol\theta, \Omega_b h^2) \,,
\end{alignat}
where $L$ is the Gaussian likelihood of the parameters $\boldsymbol{\theta}$ and $\Omega_b h^2$. The CMB contribution is
\begin{alignat}{1}
    -2 \log L_\text{CMB}(\boldsymbol\theta, \Omega_b h^2) = \left(\mathbf{x}(\boldsymbol\theta, \Omega_b h^2) - \hat{\mathbf{x}} \right)^\intercal \mathsf{C}_\text{CMB}^{-1} \left(\mathbf{x}(\boldsymbol\theta, \Omega_b h^2) - \hat{\mathbf{x}} \right) \,,
\end{alignat}
where $\mathbf{x}(\boldsymbol \theta, \Omega_b h^2) \equiv (\Omega_b h^2, N_\text{eff}(\boldsymbol\theta), \textrm{Y}_{\textrm{P}}(\boldsymbol\theta))^\intercal$, $\hat{\mathbf{x}}$ is the central value of the same parameters as reported by Planck~\cite{Aghanim:2018eyx}, and $\mathsf{C}$ is the reported covariance matrix for the three parameters we are considering.\footnote{We drop a constant term in the expression of the Gaussian likelihood, which does not affect the subsequent results.}  We use the \texttt{base\_nnu\_yhe\_plikHM\_TTTEEE\_lowl\_lowE\_post\_BAO\_lensing} dataset, which fits for both $N_\text{eff}$ and Y$_\text{P}$, in addition to the six $\Lambda$CDM parameters. For this dataset,
\begin{alignat}{1}
    \hat{\mathbf{x}} \equiv (\Omega_b h^2, N_\text{eff}, \textrm{Y}_{\textrm{P}})_\text{Planck} = (0.022358, 2.93, 0.244) \,, \qquad \mathsf{C}_\text{CMB} = \begin{bmatrix}
        3.6 \times 10^{-8} & 1.5 \times 10^{-5}  & 9.3 \times 10^{-7} \\
        1.5 \times 10^{-5} & 8.2 \times 10^{-2}  & -3.5 \times 10^{-3}\\
        9.3 \times 10^{-7} & -3.5 \times 10^{-3} & 3.2 \times 10^{-4}
    \end{bmatrix} \,.
\end{alignat}
The BBN contribution is similarly constructed as 
\begin{alignat}{1}
    -2 \log L_\text{BBN}(\boldsymbol\theta, \Omega_b h^2) = \left(\mathbf{y}(\boldsymbol\theta, \Omega_b h^2) - \hat{\mathbf{y}} \right)^\intercal \mathsf{C}_\text{BBN}^{-1}(\boldsymbol\theta, \Omega_b h^2) \left(\mathbf{y}(\boldsymbol\theta, \Omega_b h^2) - \hat{\mathbf{y}} \right) \,,
\end{alignat}
where $\mathbf{y}(\boldsymbol\theta, \Omega_b h^2) \equiv (\textrm{Y}_{\textrm{P}}(\boldsymbol\theta, \Omega_b h^2), \text{D/H}(\boldsymbol\theta, \Omega_b h^2))^\intercal$, and $\hat{\mathbf{y}}$ is the vector of observed central values of Y$_{\textrm{P}}$~\cite{Zyla:2020zbs} and D/H~\cite{Cooke:2017cwo}, given by $\hat{\mathbf{y}} = (0.245, 2.527 \times 10^{-5})^\intercal$. We use our modified version of \texttt{PRIMAT}~\cite{Pitrou:2018cgg} to compute $\mathbf{y}(\boldsymbol \theta, \Omega_b h^2)$. Unlike the CMB covariance matrix, the BBN covariance $\mathsf{C}_\text{BBN}(\boldsymbol\theta, \Omega_b h^2)$ depends on both the model parameters and baryon density. There are two contributions: \textit{i)} uncertainties in the measurement of Y$_{\textrm{P}}$ and D/H, $\sigma_{\textrm{Y}_{\textrm{P}}}^\text{ex}$ and $\sigma_\text{D/H}^\text{ex}$, which can be straightforwardly obtained from the quoted measurement uncertainties in Refs.~\cite{Zyla:2020zbs,Cooke:2017cwo}, and \textit{ii)} theoretical uncertainties in the computation of elemental abundances as a function of $\boldsymbol \theta$ and $\Omega_b h^2$ due to uncertainties in nuclear rates, which depend on $\boldsymbol\theta$ and $\Omega_b h^2$. 
\begin{figure}[t]
    \centering
    \includegraphics[width=0.85\textwidth]{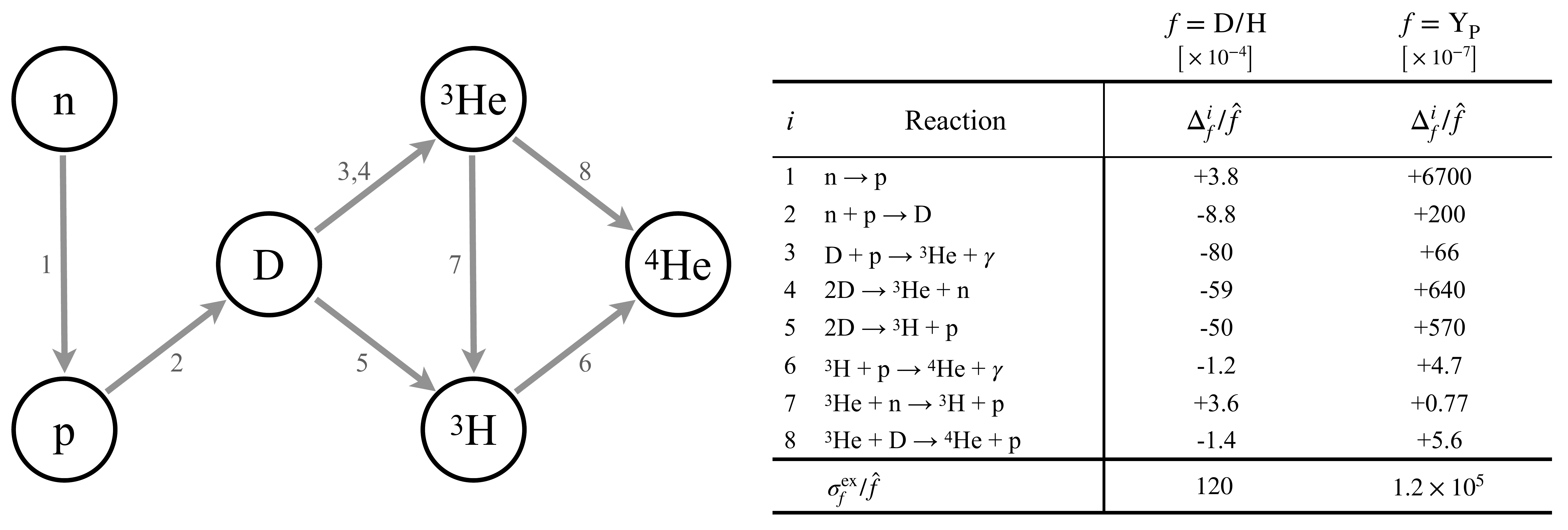}
    \caption{(Left, the {\it piscis})~The reactions in the BBN network that contribute most significantly to the theoretical uncertainties in the D/H and  Y$_{\textrm{P}}$ predictions, assuming the standard cosmological scenario ($m_{\chi}\rightarrow\infty$).  Each reaction is labeled by an index $i$ (gray number), which is specified in the associated table.  (Right)~
    The table shows the fractional change in the D/H and Y$_{\textrm{P}}$ predictions, relative to the central measured values provided in Eq.~\eqref{eq:BBN2}, that we obtain by varying the rate of reaction $i$ within $1\sigma^{\rm th}_i$ of its theoretical uncertainty.  Importantly, the theoretical uncertainty $\sigma^{\rm th}_i$ depends on both the model parameters $\boldsymbol{\theta}$ as well as $\Omega_b h^2$.  The values in this table correspond to the standard cosmological scenario with  $\Omega_b h^2 = 0.022358$, and will change for different regions of dark sector parameter space, although we find that they always remain $\lesssim 1.3\%$.  For comparison, the bottom row of the table includes the relative size of the measurement uncertainty in the D/H and Y$_{\textrm{P}}$ abundances, compared to their respective central values.  Note that the uncertainty of proton-neutron interconversion is dominated by the uncertainty in the neutron lifetime \cite{Pitrou:2018cgg}.}
    \label{fig:covariance}
\end{figure}

To estimate the theoretical uncertainties, we follow the method described in Refs.~\cite{Fiorentini:1998fv,Arbey:2016kqi} and implemented in the public BBN code \texttt{AlterBBN} \cite{Arbey:2011nf,Arbey:2018zfh} to calculate the covariances through variation of the reaction rates.  Specifically, we compare the predicted BBN abundances using the central values for each reaction rate versus the predictions when the $i^\text{th}$ reaction rate in the network is increased to the central value plus its quoted theoretical uncertainity, $\sigma^\text{th}_i(\boldsymbol\theta, \Omega_b h^2)$, which in general varies with the photon temperature.  This gives rise to a change to the elemental abundances $\Delta_f^i$, where $f$ is either Y$_{\textrm{P}}$ or D/H\@. Fig.~\ref{fig:covariance} shows the eight reactions that give the largest  shift to the fractional abundances of D/H and Y$_{\rm P}$, assuming the standard cosmological scenario with $m_\chi \rightarrow \infty$.  Because all reaction rates lead to a small ($\lesssim 1$\%) change in $\Delta_f^i/\hat{f}$, we can approximate the uncertainties in the predicted abundances as linear and add them in quadrature, i.e.,\ 
\begin{alignat}{1}
    \sigma_{\textrm{Y}_{\textrm{P}}}^\text{th} (\boldsymbol\theta, \Omega_b h^2) = \left[ \sum_i (\Delta_{\textrm{Y}_{\textrm{P}}}^i)^2 \right]^{1/2}\,, \quad \sigma_{\text{D/H}}^\text{th}(\boldsymbol \theta, \Omega_b h^2) = \left[ \sum_i (\Delta_{\text{D/H}}^i)^2 \right]^{1/2} \,, \quad \tau^\text{th}(\boldsymbol \theta, \Omega_b h^2) = \sum_i \Delta^i_{\textrm{Y}_{\textrm{P}}} \Delta^i_{\text{D/H}} \,,
\end{alignat}
where $\tau \equiv \text{cov}(\textrm{Y}_{\textrm{P}}, \text{D/H})$ is the covariance between Y$_{\textrm{P}}$ and D/H, and the sum $i$ is over the eight reactions shown in Fig.~\ref{fig:covariance}. Without any electromagnetically interacting DM and taking the central value of $\Omega_b h^2 = 0.022358$ in our Planck dataset, these quantities as computed by our modified version of \texttt{PRIMAT} are
\begin{alignat}{1}
    \sigma_{\textrm{Y}_{\textrm{P}}}^{\text{th}, \Lambda\text{CDM}} = 1.7\times 10^{-4} \quad, \quad \sigma_\text{D/H}^{\text{th}, \Lambda \text{CDM}} = 2.8\times 10^{-7} \quad, \quad \tau^\text{th}_{\Lambda\text{CDM}} = -3.0 \times 10^{-12} \, .
\end{alignat}
Note that these uncertainties are similar to those obtained using a more sophisticated Monte Carlo approach taken in the \texttt{PRIMAT} code to obtain precision BBN predictions~\cite{Pitrou:2020etk}. We emphasize that these quantities are computed for each set of model parameters $\boldsymbol \theta$ and baryon density $\Omega_b h^2$, but that $\Delta_f^i/\hat{f}$ always remains $\lesssim 1.3\%$ for both Y$_{\textrm{P}}$ and D/H, justifying the linear approximation used here.  This careful treatment of the theoretical uncertainty is important because the theoretical uncertainty for D/H is of the same order of magnitude as the measurement uncertainty, and can vary by up to 90\% with different values of $\boldsymbol \theta$ and $\Omega_b h^2$, leading to similar variations in the overall uncertainty. 

We can now add the theoretical and measured uncertainties in quadrature to obtain the full covariance matrix $\mathsf{C}_\text{BBN}$: 
\begin{alignat}{1}
     \mathsf{C}_\text{BBN}(\boldsymbol \theta, \Omega_b h^2) = \begin{bmatrix}
         (\sigma_{\textrm{Y}_{\textrm{P}}}^\text{th}(\boldsymbol \theta, \Omega_b h^2))^2 + (\sigma_{\textrm{Y}_{\textrm{P}}}^\text{ex})^2 & \tau^\text{th}(\boldsymbol \theta, \Omega_b h^2) \\
         \tau^\text{th}(\boldsymbol \theta, \Omega_b h^2) & (\sigma_\text{D/H}^\text{th}(\boldsymbol \theta, \Omega_b h^2))^2 + (\sigma_\text{D/H}^\text{ex})^2 
     \end{bmatrix} \, .\label{eq:BBN_cov}
\end{alignat}
This completes the calculation of $L(\boldsymbol \theta, \Omega_b h^2)$.

Note the theoretical BBN uncertainty on Y$_{\rm{P}}$ is treated independently from the Planck uncertainty on the same parameter.  We have verified computationally that neglecting Y$_{\rm{P}}$ entirely in the CMB constraint changes our bounds by only 5-10\%.  Therefore, any potential correlation between the CMB and BBN theoretical uncertainties on Y$_{\rm{P}}$ does not have noticeable impact on the joint constraint.

In addition to CMB constraints derived from the Planck measurement, we also include results using the forecasted covariance matrices for the upcoming Simons Observatory~\cite{SimonsObservatory:2018koc}. The relevant central values and covariance matrix are~\cite{Sabti:2019mhn}:
\begin{alignat}{2}
    \hat{\mathbf{x}}_\text{SO} &\equiv (\Omega_b h^2, N_\text{eff}, \text{Y}_\text{p})_\text{SO} = (0.022360, 3.046, 0.2472) \,,\quad \mathsf{C}_\text{CMB,SO} &&= \begin{bmatrix}
        5.3 \times 10^{-9} & 5.8 \times 10^{-7}  & 1.6 \times 10^{-7} \\
        5.8 \times 10^{-7} & 1.2 \times 10^{-2}  & -6.2 \times 10^{-4}\\
        1.6 \times 10^{-7} & -6.2 \times 10^{-4} & 4.4 \times 10^{-5}
    \end{bmatrix} \,.
\end{alignat}

\section{Modifications to Public Codes}\label{app:modifications_to_codes}

\setcounter{equation}{0}
\setcounter{figure}{0}
\setcounter{table}{0}
\renewcommand{\theequation}{C\arabic{equation}}
\renewcommand{\thefigure}{C\arabic{figure}}
\renewcommand{\thetable}{C\arabic{table}}

In this appendix, we detail the modifications made to the publicly-available versions of \texttt{nudec\_BSM}~\cite{Escudero_2019,Escudero:2020dfa} and \texttt{PRIMAT}~\cite{Pitrou:2018cgg}.  Broadly, we modify \texttt{nudec\_BSM} to calculate $T_{\gamma}$, $T_{\nu}$, the Hubble constant $H$, and the expansion rate $a$ following its normal calculation routines but including extra terms in the relevant equations for the dark photon and $\Delta N_\nu$.  The resulting thermodynamics are output to text files; these can be used on their own to calculate final values of \Neff, and can also be fed into \texttt{PRIMAT} so that the precise thermodynamics do not have to be recomputed.  We add additional modules to \texttt{PRIMAT} to read in and interpret these text files and use the results at the appropriate stage in the calculation, and finally add a new module to \texttt{PRIMAT} to calculate the linear covariance associated with any given reaction in its network.

The python code \texttt{nudec\_BSM} contains several modules for calculating \Neff\, in modified cosmologies.  To perform the calculations in this paper, we modify the script \texttt{WIMP\_e.py}, which is intended to calculate \Neff\, in the presence of a light, electrophilic WIMP with user-defined mass, statistics, and number of degrees of freedom.  We expand this module to accommodate a second particle in the dark sector (assuming the second particle is a massive vector boson, but it is straightforward to consider other scenarios of interest), as well as nonzero $\Delta N_\nu$, by modifying several sections of the code.  

Including nonzero $\Delta N_{\nu}$, a quantity defined at late times by $(\rho_\xi / \rho_\nu)_0$, involves guessing at the initial conditions of $\rho_\xi\propto a^{-4}$ at early times that yields the correct value of $\Delta N_\nu$ at the end of BBN\@. This is achieved by bracketing the target $\Delta N_\nu$ with two extreme values of $\rho_\xi$, and adopting a bisection method to find the correct initial value of $\rho_\xi$.  Therefore, this module must be run multiple times to obtain the thermodynamics for a single combination of DM parameters and nonzero $\Delta N_{\nu}$.  In order to make scans over large swaths of parameter space economical with this constraint, it is necessary to perform a number of integrals originally evaluated using the default \texttt{SciPy} numerical quadrature method in \texttt{nudec\_BSM} with a more computationally efficient method. We numerically evaluate these integrals by rewriting them as convergent infinite series, and summing over a sufficient number of leading terms, taking into account both precision and speed. This speeds up a single evaluation by roughly a factor of 10 with no significant loss in precision, while typically $\mathcal{O}(10)$ evaluations are required for a single data point, so the script can still be run reasonably quickly despite the number of evaluations required to obtain constraints. The details of our numerical integration method are left to Appendix~\ref{app:integrals}, though the series themselves will be included in the code in this appendix.

We first summarize all of the modifications made to \texttt{nudec\_BSM} briefly, and subsequently detail the modifications explicitly.  Briefly, the required modifications are:
\begin{enumerate}
    \item{Augmenting the command line input to allow the user to specify dark photon and $\Delta N_\nu > 0$ scenarios,}
    \item{Restructuring function definitions to accommodate series evaluation of Bose-Einstein and Fermi-Dirac integrals,}
    \item{Including a new flag to allow the user to toggle between the series evaluation and numerical integral evaluation,}
    \item{Implementing the series method to calculate the DM thermodynamics,}
    \item{Including functions to characterize the thermodynamics of the dark photon,}
    \item{Including functions to characterize the thermodynamics of the equivalent neutrinos,}
    \item{Updating the combined thermodynamics functions for all species to include dark photons and $\Delta N_\nu$,}
    \item{Updating the temperature evolution equations to reflect the presence of dark photons and $\Delta N_\nu$,}
    \item{Extending the period over which the thermodynamics are calculated to match the initial and final conditions of \texttt{PRIMAT},}
    \item{Implementing a method to find the initial conditions that will at late times produce the value of $\Delta N_{\nu}$ specified by the user,}
    \item{Re-running the code at high precision once these initial conditions are known,}
    \item{Calculating the scale factor, and}
    \item{Outputting the results to be read into \texttt{PRIMAT}.}
\end{enumerate}
Explicitly, these modifications take the following form:
\begin{enumerate}
    \item{We augment the command line input to include inputs for \texttt{dNnu}, which stands for $\Delta N_\nu$, and the mass ratio $m_{A'}/m_\chi$, given in the script by \texttt{ratio}.  The dark photon mass \texttt{MDP} is set using this second input.}
    \item{In addition to importing additional packages in the preamble required later for computing sums or integrals, we also include in the beginning of the code all-purpose functions for computing the energy densities and pressures of general particle species:
    \begin{lstlisting}[language=Python]
  from scipy.integrate import trapz
  from scipy.special import kv
  
  # Define Integrands
    def rho_integrand(E, m_over_T, SPIN):
        if SPIN == 'B' or SPIN == 'BOSE' or SPIN == 'Bose':
            return E**2*(E**2-(m_over_T)**2)**0.5/(np.exp(E)-1.)
        elif SPIN == 'F' or SPIN == 'Fermi' or SPIN == 'Fermion': 
            return E**2*(E**2-(m_over_T)**2)**0.5/(np.exp(E)+1.)
        else: 
            raise TypeError('Invalid SPIN.')
    
    def p_integrand(E, m_over_T, SPIN): 
        if SPIN == 'B' or SPIN == 'BOSE' or SPIN == 'Bose': 
            return (E**2-(m_over_T)**2)**1.5/(np.exp(E)-1.)
        elif SPIN == 'F' or SPIN == 'Fermi' or SPIN == 'Fermion': 
            return (E**2-(m_over_T)**2)**1.5/(np.exp(E)+1.)
        else: 
            raise TypeError('Invalid SPIN.')
    
    def drho_dT_integrand(E, m_over_T, SPIN): 
        if SPIN == 'B' or SPIN == 'BOSE' or SPIN == 'Bose': 
            return 0.25*E**3*(E**2-(m_over_T)**2)**0.5*np.sinh(E/2.0)**-2
        elif SPIN == 'F' or SPIN == 'Fermi' or SPIN == 'Fermion': 
            return 0.25*E**3*(E**2-(m_over_T)**2)**0.5*np.cosh(E/2.0)**-2
    
    def int_3_2(a, SPIN): 
        k_plus_1_vec = np.arange(20) + 1
        K_terms = kv(2, a*(k_plus_1_vec))
        if SPIN == 'B' or SPIN == 'BOSE' or SPIN == 'Bose':    
            b = 1
        elif SPIN == 'F' or SPIN == 'Fermi' or SPIN == 'Fermion':     
            b = -1
        return 3 * a**2 *  np.sum((b**(k_plus_1_vec-1)) * K_terms / k_plus_1_vec**2)
    
    def int_1_2(a, SPIN): 
        k_plus_1_vec = np.arange(20) + 1
        K_terms = kv(1, a*(k_plus_1_vec)) 
        if SPIN == 'B' or SPIN == 'BOSE' or SPIN == 'Bose':   
            b = 1 
        elif SPIN == 'F' or SPIN == 'Fermi' or SPIN == 'Fermion':     
            b = -1
        return a * np.sum(b**(k_plus_1_vec-1) * K_terms / k_plus_1_vec)
    
    def int_drho_dT(a, SPIN): 
        k_plus_1_vec = np.arange(20) + 1
        K_3_terms = kv(3, a*(k_plus_1_vec))
        K_2_terms = kv(2, a*(k_plus_1_vec))
    
        if SPIN == 'B' or SPIN == 'BOSE' or SPIN == 'Bose':    
            b = 1
        elif SPIN == 'F' or SPIN == 'Fermi' or SPIN == 'Fermion':    
            b = -1
            
        series_1 = 3 * a**3 * (b**(k_plus_1_vec-1) * K_3_terms / k_plus_1_vec)
        series_2 = a**4 * (b**(k_plus_1_vec-1) * K_2_terms)
        return np.sum(series_1 + series_2)
    \end{lstlisting}}
    \item{We include an option \texttt{seriesFlag}, which is set to \texttt{True} for series evaluation of numerical integrals, or \texttt{False} for numerical evaluation of these integrals.}
    \item{We augment the functions in the \texttt{WIMP Thermodynamics} section to evaluate the numerical integrals using the series method described in Appendix~\ref{app:integrals} if \texttt{seriesFlag} is set to \texttt{True}:
    \begin{lstlisting}[language=Python]
  # WIMP Thermodynamics
    def rho_DM(T, MDM, series=seriesFlag): 
        if T < MDM/30.0:         
            return 0.0 
        else:
            if series: 
                if MDM/T < 0.01: 
                    if SPIN == 'F' or SPIN == 'Fermi' or SPIN == 'Fermion': 
                        fac = 7. / 8. * np.pi**4 / 15. 
                    elif SPIN == 'B' or SPIN == 'BOSE' or SPIN == 'Bose': 
                        fac = np.pi ** 4 / 15. 
                    return gDM / (2* np.pi**2) * T**4 * fac            
                else:
                    return gDM / (2* np.pi**2) * T**4 * (
                        int_3_2(MDM/T, SPIN) + (MDM/T)**2 * int_1_2(MDM/T, SPIN)
                    )
            else:
                return gDM / (2 * np.pi**2) * T**4 * quad(
                    rho_integrand, MDM/T, 100, 
                    args=(MDM/T, SPIN), epsabs=1e-8, epsrel=1e-8
                )[0]
    
    def p_DM(T, MDM, series=seriesFlag): 
        if T < MDM/30.0: 
            return 0.0     
        else: 
            if series: 
                if MDM/T < 0.01: 
                    if SPIN == 'F' or SPIN == 'Fermi' or SPIN == 'Fermion': 
                        fac = 7. / 8. * np.pi**4 / 15. 
                    elif SPIN == 'B' or SPIN == 'BOSE' or SPIN == 'Bose': 
                        fac = np.pi ** 4 / 15. 
                    return gDM / (6 * np.pi**2) * T**4 * fac             
                else:
                    return gDM / (6 * np.pi**2) * T**4 * int_3_2(MDM/T, SPIN)
            else:
                return gDM / (6 * np.pi**2) * T**4 * quad(
                    p_integrand, MDM/T, 100, args=(MDM/T, SPIN), epsabs=1e-8, epsrel=1e-8
                )[0]
    
    def drho_DMdT(T, MDM, series=seriesFlag): 
        if T < MDM / 30.0: 
            return 0.0 
        else: 
            if series: 
                if MDM/T < 0.01: 
                    if SPIN == 'F' or SPIN == 'Fermi' or SPIN == 'Fermion': 
                        fac = 7 * np.pi**4 / 30
                    elif SPIN == 'B' or SPIN == 'BOSE' or SPIN == 'Bose': 
                        fac = 4 * np.pi**4 / 15
                    return gDM / (2 * np.pi**2) * T**3 * fac
                else: 
                    return gDM / (2 * np.pi**2) * T**3 * int_drho_dT(MDM/T, SPIN)
            else:
                return gDM / (2 * np.pi**2) * T**3 * quad(
                    drho_dT_integrand, MDM/T, 100,
                    args=(MDM/T, SPIN), epsabs=1e-8, epsrel=1e-8
                )[0]
	\end{lstlisting}
    Note these quantities are negligible by the time the temperature of the species drops below $1/30$ of the species mass, and are therefore set to zero in order to avoid numerical overflow. We also use the relativistic expressions for all quantities whenever the particle mass is less than 1\% of its temperature, where this is an excellent approximation.}
	\item{We also add similar, new functions to characterize the thermodynamics of the dark photon in the same section:
	\begin{lstlisting}[language=Python]
	  def p_DM_DP(T, MDP, series=seriesFlag): 
    if T < MDP / 30.0:          
        return 0.0        
    else: 
        if series: 
            if MDP/T < 0.01: 
                fac = np.pi ** 4 / 15. 
                return 3. / (6 * np.pi**2) * T**4 * fac             
            else:
                return 3. / (6 * np.pi**2) * T**4 * int_3_2(MDP/T, 'B')
        else:        
            return 3. / (6 * np.pi**2) * T**4 * quad(
                p_integrand, MDP/T, 100, args=(MDP/T, 'B'), epsabs=1e-8, epsrel=1e-8
            )[0]

    def drho_DM_DPdT(T, MDP, series=seriesFlag):
        if T < MDP/30.0:   
            return 0.0       
        else: 
            if series: 
                if MDP/T < 0.01: 
                    fac = 4 * np.pi**4 / 15
                    return 3. / (2 * np.pi**2) * T**3 * fac
                else: 
                    return 3. / (2 * np.pi**2) * T**3 * int_drho_dT(MDP/T, 'B')
            else:       
                return 3. / (2 * np.pi**2) * T**3 * quad(
                    drho_dT_integrand, MDP/T, 100, args=(MDP/T, 'B'), epsabs=1e-8, epsrel=1e-8
                )[0]
	\end{lstlisting}
	}
	\item{We modify the \texttt{Thermodynamics} and \texttt{Derivatives} sections to include functions to characterize the equivalent neutrino energy density.  We also modify the functions for electrons in these sections to use the series method as well if \texttt{seriesFlag=True}.
	\begin{lstlisting}[language=Python]
  # Thermodynamics
  \end{lstlisting}
  \qquad\vdots
  \begin{lstlisting}[language=Python]
    def rho_eqnu(T): return 2 * 7./8. * np.pi**2/30. * T**4
    def rho_e(T, series=seriesFlag): 
        if T < me / 30.0: 
            return 0.0 
        else:
            if series: 
                if me/T < 0.01: 
                    fac = 7. / 8. * np.pi**4 / 15.  
                    return 4. / (2* np.pi**2) * T**4 * fac            
                else:
                    return 4. / (2* np.pi**2) * T**4 * (
                        int_3_2(me/T, 'F') + (me/T)**2 * int_1_2(me/T, 'F')
                    )
            else: 
                return 4. / (2 * np.pi**2) * T**4 * quad(
                    rho_integrand, me/T, 100, args=(me/T, 'F'), epsabs=1e-12, epsrel=1e-12
                )[0]
    
    def p_e(T, series=seriesFlag): 
        if T < me / 30.0: 
            return 0.0 
        else:
            if series: 
                if me/T < 0.01: 
                    fac = 7. / 8. * np.pi**4 / 15.  
                    return 4. / (6 * np.pi**2) * T**4 * fac             
                else:
                    return 4. / (6 * np.pi**2) * T**4 * int_3_2(me/T, 'F')
            else: 
                return 4. / (6 * np.pi**2) * T**4 * quad(
                    p_integrand, me/T, 100, args=(me/T, 'F'), epsabs=1e-12, epsrel=1e-12
                )[0]
    
  # Derivatives
    \end{lstlisting}
    \qquad\vdots
    \begin{lstlisting}[language=Python]
    def drho_eqnudT(T):   return 4*rho_eqnu(T)/T
    def drho_edT(T, series=seriesFlag):
        if T < me/30.0:         
            return 0.0
        if series: 
            if me/T < 0.01: 
                fac = 7 * np.pi**4 / 30
                return 4. / (2 * np.pi**2) * T**3 * fac
            else: 
                return 4. / (2 * np.pi**2) * T**3 * int_drho_dT(me/T, 'F')
        else:             
            return 4./(2*np.pi**2)*T**3 * quad(
                drho_dT_integrand, me/T, 100, args=(me/T, 'F'), epsabs=1e-8, epsrel=1e-8
            )[0]
	\end{lstlisting}
	}
	\item{We define the total energy density, pressure, and Hubble rate to include contributions from the second particle in the dark sector and $\Delta N_\nu$:
	\begin{lstlisting}[language=Python]
  # Add contributions from dark photon and equivalent neutrino 
  # to total energy density, total pressure and Hubble rate

  # Rho tot; neglects the possibility of neutrino asymmetry
    def rho_tot(T_gam,T_nue,T_numu,Tnu_eq,MDM): 
        return (
            rho_gam(T_gam) + rho_e(T_gam) + rho_DM(T_gam,MDM) 
            + rho_DM_DP(T_gam,ratio*MDM) + rho_nu(T_nue) + 2*rho_nu(T_numu) 
            + rho_eqnu(Tnu_eq) - P_int(T_gam) + T_gam*dP_intdT(T_gam)
        )
  # P tot
    def p_tot(T_gam,T_nue,T_numu,Tnu_eq,MDM):
        return  (
            1./3. * rho_gam(T_gam) + p_e(T_gam) + p_DM(T_gam,MDM) 
            + p_DM_DP(T_gam,ratio*MDM) + 1./3. * rho_nu(T_nue) + 1./3. * 2*rho_nu(T_numu) 
            + 1./3. * rho_eqnu(Tnu_eq) + P_int(T_gam)
        )
  # Hubble
    def Hubble(T_gam,T_nue,T_numu,Tnu_eq,MDM):
        return FAC * (rho_tot(T_gam,T_nue,T_numu,Tnu_eq,MDM)*8*np.pi/(3*Mpl**2))**0.5
	\end{lstlisting}}
	\item{We modify functions in the \texttt{Temperature Evolution Equations} section to include these new contributions, and add a new function to calculate the scale factor as a function of time:
	\begin{lstlisting}[language=Python]
  # Update photon temperature and neutrino temperature derivative functions 
  # to include dark photon and equivalent neutrino contributions.
  # Also, include a function to calculate the scale factor.

    def dTnu_dt(T_gam,T_nue,T_numu,Tnu_eq,MDM):
        return -(
            Hubble(T_gam,T_nue,T_numu,Tnu_eq,MDM) * (3 * 4 * rho_nu(T_nue) ) 
            - (2*DeltaRho_numu(T_gam,T_nue,T_numu) + DeltaRho_nue(T_gam,T_nue,T_numu))
        ) / (3*drho_nudT(T_nue))
    
    def dTgam_dt(T_gam,T_nue,T_numu,Tnu_eq,MDM):
        return -(
            Hubble(T_gam,T_nue,T_numu,Tnu_eq,MDM)*( 4*rho_gam(T_gam) 
            + 3*(rho_e(T_gam)+p_e(T_gam)) + 3*(rho_DM(T_gam,MDM)+p_DM(T_gam,MDM) 
            + rho_DM_DP(T_gam,ratio*MDM) + p_DM_DP(T_gam,ratio*MDM)) 
            + 3 * T_gam * dP_intdT(T_gam)) + DeltaRho_nue(T_gam,T_nue,T_numu) 
            + 2*DeltaRho_numu(T_gam,T_nue,T_numu) 
        )/ ( 
            drho_gamdT(T_gam) + drho_edT(T_gam) + drho_DMdT(T_gam,MDM) 
            + drho_DM_DPdT(T_gam,ratio*MDM) + T_gam * d2P_intdT2(T_gam) 
        )
    
    def dTeqnu_dt(T_gam,T_nue,T_numu,Tnu_eq,MDM):
        return -(
            Hubble(T_gam,T_nue,T_numu,Tnu_eq,MDM) * (4*rho_eqnu(Tnu_eq))
        )/drho_eqnudT(Tnu_eq)
    
    def dT_totdt(vec,t,MDM):
        T_gam, T_nu, Tnu_eq = vec
        return [
            dTgam_dt(T_gam,T_nu,T_nu,Tnu_eq,MDM),
            dTnu_dt(T_gam,T_nu,T_nu,Tnu_eq,MDM),
            dTeqnu_dt(T_gam,T_nu,T_nu,Tnu_eq,MDM)
        ]
 
  # Find a(t)
    def scale_fac(i,a0):
        log_a_over_a0 = trapz(HubbleTime[:i+1],tvec[:i+1])
        return np.exp(log_a_over_a0)* a0
	\end{lstlisting}
	(The array \texttt{HubbleTime} is defined later in the code.)}
	\item{We modify the initial and final times and temperatures for the integration so the output may be passed to \texttt{PRIMAT} without any extrapolation.  We begin the calculation at a common temperature of $\texttt{T0}=\SI{100}{\mega\eV}$ and increase the end time \texttt{t\_max} to $5\times 10^{6}$ seconds.}
	\item{We implement a guess-and-check method to find the correct initial temperature of the inert species that leads to the user-specified value of $\Delta N_{\nu}$.  This involves using a bisection method, where a bounded region that includes the correct initial temperature is identified, and then initial conditions are chosen at the midpoint of that region.  The size of the bounding region is halved depending on whether the resulting $\Delta N_{\nu}$ is too high or too low, and the initial temperature is chosen at the new midpoint.  This process repeats until the correct value of $\Delta N_{\nu}$ is achieved.  This block also contains several lines to ensure the tolerance of the integrator \texttt{odeint} is reduced appropriately as the bounded region decreases in size.
	\begin{lstlisting}[language=Python]
  # Guess-and-check the inert species initial temperature until the target 
  # value of dNnu (at late times) is obtained, using a bisection method.
  
    dNnuTarget = round(dNnuTarget,3)
    dNnuFacMin = dNnuTarget**0.25
    dNnuFacMax = 1.2
    
    dNnu = 'placeholder'
    bounded = False
    dNnuFac = dNnuFacMax
    rtolerance = 1e-4
    rtolermax = rtolerance
    iter = 0
    
    
    while dNnu != dNnuTarget:
    	
        if dNnuTarget == 0:
            dNnuFac = 1e-6  # sufficiently small so as to reproduce 0
    
        # Start the integration at a common temperature of T0\approx 86 MeV (100) to include 
        # PRIMAT's Ti
        T0 = 100.0
        t0 = 1./(2*Hubble(T0,T0,T0,dNnuFac*T0,MDM))
    
        # Finish the calculation at t = 5e6 seconds to accommodate PRIMAT
        t_max = 5e6
    
        # Calculate
        tvec = np.logspace(np.log10(t0),np.log10(t_max),num=100)
        
        toleranceMultiple = 0.1
        while True:
    		if dNnuFacMax - dNnuFacMin > 0:
    			rtolerance = np.min((0.1 * (dNnuFacMax - dNnuFacMin) / dNnuFac, rtolermax))
    		else:
    			rtolerance = rtolermax
    			
    		try:
    			sol = odeint(dT_totdt, [T0,T0,dNnuFac*T0], tvec, args=(MDM,), rtol = rtolerance)
    			break
    		except:
    			rtolerance *= toleranceMultiple
    
        dNnu = round(rho_eqnu(sol[-1,2])/rho_nu(sol[-1,1]),3)
        
        if dNnuTarget == 0: 
            break
    
        if dNnu < dNnuTarget:
            if not bounded:  
                dNnuFacMax *= 1.2
                dNnuFac = dNnuFacMax 
            else:
                dNnuFacMin = dNnuFac
                dNnuFac = (dNnuFacMin + dNnuFacMax) / 2.
                
        elif dNnu > dNnuTarget:
            bounded = True
            dNnuFacMax = dNnuFac
            dNnuFac = (dNnuFacMin + dNnuFacMax) / 2. 
    
        else:
            break 
        
        iter += 1
        if iter >= 20: # start over with a lower tolerance, if the solver gets stuck
    		dNnuFacMin = dNnuTarget**0.25
    		dNnuFacMax = 1.2
    		dNnu = 'placeholder'
    		bounded = False
    		dNnuFac = dNnuFacMax
    		rtolerance = rtolermax*0.1
    		rtolermax *= 0.1
    		iter = 0
	\end{lstlisting}}
	\item{After the correct value of $\Delta N_{\nu}$ is obtained, the thermodynamics are computed once more at greater precision to be output and ultimately read by \texttt{PRIMAT}.  We set the tolerance of \texttt{odeint} to $10^{-8}$ and the number of points in time at which the system is evaluated to 6000.}
	\item{After the thermodynamic quantities are calculated in the code body, we calculate the Hubble rate and the scale factor using the output \texttt{tvec} and \texttt{sol} from the \texttt{Calculate} section:
	\begin{lstlisting}[language=Python]
  # Calculate thermodynamic quantities and the scale factor as 
  # functions of time and store them so that they may be output

  # Calculate the relevant Thermodynamic quantities
    rho_vec, p_vec = np.zeros(len(tvec)),np.zeros(len(tvec))
    for i in range(len(tvec)):
        rho_vec[i], p_vec[i]  = (
            rho_tot(sol[i,0], sol[i,1], sol[i,1],sol[i,2],MDM), 
            p_tot(sol[i,0], sol[i,1], sol[i,1],sol[i,2],MDM)
        )
    
  # Find the scale factor    
    HubbleTime=np.zeros(len(tvec))
    for i in range(len(tvec)):
        HubbleTime[i] = Hubble(sol[i,0],sol[i,1],sol[i,1],sol[i,2],MDM)
    Ht=interp1d(tvec,HubbleTime)
    Tgam=sol[:,0]
    final_a = 2.3487596e-10 / Tgam[-1]
    log_final_a_over_a0 = trapz(HubbleTime,tvec)
    a0 = 1./np.exp(log_final_a_over_a0) * final_a
    
    a=np.zeros(len(tvec))
    for i in range(len(tvec)):
        a[i]=scale_fac(i,a0)
	\end{lstlisting}
	}
	\item{Finally, we augment the output to include a column for Hubble and the scale factor as functions of time.}
\end{enumerate}

\lstset{style=mathstyle}

\par
Once the thermodynamics calculation is updated for these scenarios as described above, the resulting output is piped into the Mathematica code \texttt{PRIMAT}.  We add a function to \texttt{PRIMAT} to load in and store these results.  While the newest version of \texttt{PRIMAT} contains routines for precisely calculating corrections from QED and non-instantaneous neutrino decoupling, similar to \texttt{nudec\_BSM}, it is simpler and less computationally intensive to use the results from \texttt{nudec\_BSM} in the BBN calculation rather than modifying \texttt{PRIMAT} to accommodate alternative cosmologies on its own and recomputing the relevant thermodynamics.  We therefore eliminate the sections of \texttt{PRIMAT} responsible for calculating thermodynamics; this includes all sections under the header \texttt{Thermodynamics of the plasma} except for \texttt{Scale factor determination}, and the \texttt{Friedmann equation} section under the header \texttt{Time integration of Cosmology and BBN}.  Several modifications are included in place of these sections.  We also add a linear covariance estimation routine to \texttt{PRIMAT}, similar to the one used in \texttt{AlterBBN}~\cite{Arbey:2011nf,Arbey:2018zfh}.  Again, we first briefly summarize these modifications and then detail them explicitly below: 
\begin{enumerate}
    \item {Including an option to calculate abundances with central$+1\sigma$ values for one specified reaction rate in order to estimate the covariance,}
    \item{Including options to characterize the additional DM, dark photons, and $\Delta N_\nu$,}
    \item{Creating a function to load the specified precomputed thermodynamics from \texttt{nudec\_BSM} output,}
    \item{Eliminating or replacing the functions in \texttt{PRIMAT} that compute the same thermodynamics as those loaded from \texttt{nudec\_BSM} output,}
    \item{Updating the naming conventions for additional thermodynamics computed by \texttt{PRIMAT} to reflect the characteristics of the alternative cosmology,}
    \item{Adding a linear covariance estimation routine into the appropriate functions in \texttt{PRIMAT}, and}
    \item{Redefining the functions used to call \texttt{PRIMAT}'s abundance calculation routines to reflect these changes.}
\end{enumerate}

In more detail, these modifications take the following form:
\begin{enumerate}
    \item{Under \texttt{Preambule} $\rightarrow$ \texttt{Options} $\rightarrow$ \texttt{Numerical options}, we add a flag \texttt{\$LinearCovarianceEstimation}; when set to ``True", \texttt{PRIMAT} will calculate BBN abundances using the central$+ 1\sigma$ value for a given reaction rate specified later in the code.}
    \item {We include another section after the \texttt{Initial Definitions} section that is titled \texttt{Alternative Cosmology Parameters}.  This section includes definitions of $\Delta N_\nu$ (\texttt{dNnu}), strings indicating the DM mass and number of degrees of freedom \texttt{mchiStr} and \texttt{gchi}, respectively, a flag \texttt{\$DarkPhotonMediator} to specify whether a dark photon mediator is present in the dark sector, and a string \texttt{ratioStr} to specify the mass ratio $m_{A'}/m_{\chi}$ if the dark photon is present.  There are also flags \texttt{\$DMeBoson} and \texttt{\$DMeFermion} that the user can use to specify whether the DM is bosonic or fermionic.} 
    \item{We replace most of the sections under the \texttt{Thermodynamics of the plasma} header by a single section, \texttt{Thermodynamics from nudec\_BSM}.  Here, the output from the previous section is imported and processed by \texttt{PRIMAT} (the user should define \texttt{nudecBSMFilePath} appropriately):
    \begin{lstlisting}
  (* Define a function to load in the precomputed thermodynamics from nudec_BSM\@. 
  Then, call that function. *)

  nudecBSMFilePath = 
  
  LoadNudecBSM := (

      DPPrefix = If[$DarkPhotonMediator, "DP_", ""]; 
      SpinChar = Which[$DMeBoson, "B", $DMeFermion, "F"]; 
      filename = StringTemplate["mchi-`a`_g-`b`_Spin-`c`_dNnu-`d`_mratio-`e`.dat"][
          <|"a" -> mchiStr, "b" -> gchi, "c" -> SpinChar, "d" -> dNnu, "e" -> ratioStr|>
      ];
      nudecBSMThermo = Import[
          nudecBSMFilePath <> "/Results/WIMPS/" <> DPPrefix <> filename
      ];
        
      (*Functions of Plasma Temperature*)
        
      T\[Nu]OverTgammaOFTgamma = Join[
          Transpose[List[nudecBSMThermo[[All, 2]]]], Transpose[List[nudecBSMThermo[[All, 3]]]], 2
      ];
      aOFTgamma = Join[
          Transpose[List[nudecBSMThermo[[All, 2]]]], Transpose[List[nudecBSMThermo[[All, 5]]]], 2
      ]; 

      T\[Nu]overT[Tv_] := Interpolation[T\[Nu]OverTgammaOFTgamma][Tv];
      a[Tv_] := Interpolation[aOFTgamma][Tv];
        
      (* Functions of Time *);
        
      TgammaOFTime = Join[
          Transpose[List[nudecBSMThermo[[All, 1]]]], Transpose[List[nudecBSMThermo[[All, 2]]]], 2
      ];
        
      aOFTime = Join[
          Transpose[List[nudecBSMThermo[[All, 1]]]], Transpose[List[nudecBSMThermo[[All, 5]]]], 2
      ]; 
        
      Toft[t_] := Interpolation[TgammaOFTime][t];
      aoft[t_] := Interpolation[aOFTime][t];
      
      (* Functions of Scale Factor *);
        
      TimeOFa = Join[
          Transpose[List[nudecBSMThermo[[All, 5]]]], Transpose[List[nudecBSMThermo[[All, 1]]]], 2
      ]; 
        
      TgammaOFa = Join[
          Transpose[List[nudecBSMThermo[[All, 5]]]], Transpose[List[nudecBSMThermo[[All, 2]]]], 2
      ];
        
      HubbleOFa = Join[
          Transpose[List[nudecBSMThermo[[All, 5]]]], Transpose[List[nudecBSMThermo[[All, 4]]]], 2
      ];
        
      tofa[a_] := Interpolation[TimeOFa][a];
      Tofa[a_] := Interpolation[TgammaOFa][a];
      H[a_] := Interpolation[HubbleOFa][a];
  );
    
  LoadNudecBSM;
    \end{lstlisting}}
    \item{We remove the definitions of the functions \texttt{a[T]} and \texttt{Tofa[a]} from the remaining section under this header, \texttt{Scale factor determination}.  Similarly, we eliminate the definitions of \texttt{tofa[a]}, \texttt{aoft[t]}, and \texttt{Toft[t]} from the section \texttt{Time integration of Cosmology and BBN} $\rightarrow$ \texttt{Time and scale factor} as they are now defined above.}
    \item{We modify the \texttt{Weak reactions n+$\nu$ $\leftrightarrow$ p+e} $\rightarrow$ \texttt{Precomputation and storage of rates} section such that new files encoding the weak rates are created and stored when DM is included.  It is necessary to recompute the weak rates in the presence of DM since the rate of proton-neutron interconversion is sensitive to the photon and neutrino temperatures, which are in turn affected by the presence of DM\@.  We add new blocks to redefine the file naming convention:
    \begin{lstlisting}
  (* Change the weak rate file naming conventions to reflect whether there was a dark photon present 
  and the mass and characteristics of the DM in the scenario in which the weak rates were computed *)
  
  DPPrefix = If[$DarkPhotonMediator, "DP_", ""];
  SpinChar = Which[$DMeBoson, "B", $DMeFermion, "F"];
  If[
      ratioStr == "3.0" || ratioStr == 3.0,
      WRfilename = StringTemplate["mchi-`a`_g-`b`_Spin-`c`_dNnu-`d`_Omegab-`e`"][
          <|"a" -> mchiStr, "b" -> gchi, "c" -> SpinChar, "d" -> dNnu, "e" -> h2\[CapitalOmega]b0|>
      ], 
      WRfilename = StringTemplate["mchi-`a`_g-`b`_Spin-`c`_dNnu-`d`_Omegab-`e`_mratio-`f`"][
          <|"a" -> mchiStr, "b" -> gchi, "c" -> SpinChar, "d" -> dNnu, "e" -> h2\[CapitalOmega]b0, 
          "f" -> ratioStr|>
      ]
  ];
    \end{lstlisting}
    }
    This change is also reflected in the definition of the function \texttt{RateImport} in the section:
    \begin{lstlisting}
  (* Tell PRIMAT to look for precomputed rates with the file naming convention defined above.  
  If those files are not found, recompute weak rates. *)

  RateImport := (
      DPPrefix = If[$DarkPhotonMediator, "DP_", ""];
      SpinChar = Which[$DMeBoson, "B", $DMeFermion, "F"];
      If[ratioStr == "3.0" || ratioStr == 3.0, 
          WRfilename = StringTemplate["mchi-`a`_g-`b`_Spin-`c`_dNnu-`d`_Omegab-`e`"][
              <|"a" -> mchiStr, "b" -> gchi, "c" -> SpinChar, "d" -> dNnu, "e" -> h2\[CapitalOmega]b0|>
          ], 
          WRfilename = StringTemplate["mchi-`a`_g-`b`_Spin-`c`_dNnu-`d`_Omegab-`e`_mratio-`f`"][
              <|"a" -> mchiStr, "b" -> gchi, "c" -> SpinChar, "d" -> dNnu, "e" -> h2\[CapitalOmega]b0, 
              "f" -> ratioStr|>
          ]
      ]; 

      NamePENFilenp = "Interpolations/PENRatenp" <> DPPrefix <> WRfilename <> ".dat";
      NamePENFilepn = "Interpolations/PENRatepn" <> DPPrefix <> WRfilename <> ".dat";
        
      TabRatenp = Check[
          Import[NamePENFilenp, "TSV"], 
          Print["Precomputed n -> p rate not found. We recompute the rates and store them. This can take very long"]; 
          $Failed, Import::nffil
      ];
        
      TabRatepn = Check[
          Import[NamePENFilepn, "TSV"], 
          Print["Precomputed p -> n rate not found. We recompute the rates and store them. This can take very long"]; 
          $Failed, Import::nffil
      ];
        
      Timing[
          If[
              TabRatenp === $Failed || TabRatepn === $Failed || $RecomputeWeakRates,
              PreComputeWeakRates;
              Export[NamePENFilenp, TabRatenp, "TSV"];
              Export[NamePENFilepn, TabRatepn, "TSV"];,
              \[Lambda]nTOpI = MyInterpolationRate[ToExpression[TabRatenp]];
              \[Lambda]pTOnI = MyInterpolationRate[ToExpression[TabRatepn]];
          ];
      ]
  )
    \end{lstlisting}
    \item{The following items encompass all of the changes made to the section \texttt{Nuclear reactions network} and most of the changes required to add the linear covariance estimation routine to \texttt{PRIMAT}.  Note these modifications only make it possible to compute the BBN abundances using the central$+1\sigma$ value for one specified reaction rate in the reaction network; the user still must write a separate script that calls this routine, modifying the reactions in the network whose uncertainties contribute most to the uncertainties of the desired abundances (see Appendix~\ref{app:joint_constraint_details}) and computes the covariances if they wish to access the theoretical uncertainties for the abundance predictions.
    \begin{enumerate}
        \item{We modify the function \texttt{TreatReactionLine} in the subsection \texttt{Nuclear Reaction rates} $\rightarrow$ \texttt{Importation of reactions from external files (336 reactions)}:
        \begin{lstlisting}
  (* Include a condition to import the modified reaction rate for a specified reaction 
  (indicated later in the code) if the user sets the $LinearCovarianceEstimation flag to True *)
  TreatReactionLine[line_] := 
        \end{lstlisting}
        \qquad \qquad \ldots
        \begin{lstlisting}
      table = Map[{Giga #[[1]], #[[2]] Hz, #[[3]]} &, data];
      Tmin = Last[table][[1]];
      rmin = Last[table][[2]];
      Lname = ToExpression["Hold@L" <> Name];
      rv = NormalRealisation;
      MySet[Lname, MyInterpolationRate[
          {#[[1]], Identity[
              rescalefactor #[[2]] * Which[
                  $RandomNuclearRates, TruncateRateVariation[#[[3]]^rv], 
                  $LinearCovarianceEstimation, If[
                      ToString[Name] == ToString[ChosenName], 
                      TruncateRateVariation[#[[3]]], 1
                  ], 
                  True, 1
              ]
          ]} & /@ table
      ]];
  
      (* We do not rescale the reverse because it is computed FROM the forward rate. 
      So rescaling the forward rate by rescalefactor rescales them both *)
            
      ReverseReaction[Name, FrontFactor, PoweronT9, Qoverkb];
      {Name, InitialParticles, FinalParticles, rv, ReferencePaper}];
        \end{lstlisting}
        }
        \item{In the subsection \texttt{Nuclear Reaction rates} $\rightarrow$ \texttt{Collecting all reaction rates}, we add an array to easily access the names used to index the reactions in the network:
        \begin{lstlisting}
  (* An array of abbreviations of all of the reactions in the PRIMAT reaction network, 
  which allows the user to more easily specify which reaction rate they wish to adjust 
  to find the linear covariance *)
  NamesArray = {"nTOp", "npTOdg", "dpTOHe3g", "ddTOHe3n", "ddTOtp", 
      "tpTOag", "tdTOan", "taTOLi7g", "He3nTOtp", "He3dTOap", 
      "He3aTOBe7g", "Be7nTOLi7p", "Li7pTOaa", "Li7pTOaag", "Be7nTOaa", 
      "daTOLi6g", "Li6pTOBe7g", "Li6pTOHe3a", "Be9tTOB11n", "O18nTOO19g",
      "Li9pTOHe6a", "Li9dTOBe10n", "Be10aTOC14g", "N12nTOC12p", 
      "Li9pTOBe9n", "Li9aTOB12n", "Li9pTOBe10g", "N13nTON14g", 
      "B10aTON14g", "B8aTON12g", "B12pTOBe9a", "Be10pTOB11g", 
      "Be10pTOLi7a", "Be11pTOLi8a", "Be11pTOB11n", "B8nTOaap", 
      "B10nTOB11g", "B10aTOC13p", "O17nTOO18g", "F17nTOO17p", 
      "F18nTOO18p", "Be10aTOC13n", "Be11aTOC14n", "N14aTOF18g", 
      "N15aTOF19g", "O15aTONe19g", "O16pTOF17g", "O16aTONe20g", 
      "O17pTOF18g", "O18pTOF19g", "O18aTONe22g", "F17pTONe18g", 
      "F18pTONe19g", "Ne19pTONa20g", "O17pTON14a", "O18pTON15a", 
      "F18pTOO15a", "C14aTOO18g", "C14pTON15g", "Be12pTOLi9a", 
      "Li6He3TOaap", "Li6tTOBe9g", "Li6tTOaan", "Li6tTOLi8p", 
      "Li7dTOBe9g", "Li7He3TOB10g", "Li7He3TOLi6a", "Li7tTOBe10g", 
      "Li8aTOB12g", "Li8aTOB11n", "Li8dTOBe10g", "Li8He3TOB11g", 
      "Li8He3TOB10n", "Li8He3TOBe10p", "Li8He3TOLi7a", "Li8tTOBe11g", 
      "Li8tTOBe10n", "Li9aTOB13g", "Li9dTOBe11g", "Li9He3TOB12g", 
      "Li9He3TOB11n", "Li9He3TOBe11p", "Li9He3TOLi8a", "Li9tTOBe12g", 
      "Li9tTOBe11n", "Be7He3TOC10g", "Be7tTOB10g", "Be7tTOBe9p", 
      "Be7tTOLi6a", "Be9aTOC13g", "Be9dTOB11g", "Be9dTOB10n", 
      "Be9dTOBe10p", "Be9dTOLi7a", "Be9He3TOC12g", "Be9He3TOC11n", 
      "Be9He3TOB11p", "Be9He3TOaaa", "Be9tTOB12g", "Be9tTOLi8a", 
      "Be10dTOB12g", "Be10dTOB11n", "Be10dTOLi8a", "Be10He3TOC13g", 
      "Be10He3TOC12n", "Be10He3TOB12p", "Be10He3TOBe9a", "Be10tTOB13g", 
      "Be10tTOB12n", "Be10tTOLi9a", "Be11aTOC15g", "Be11dTOB13g", 
      "Be11dTOB12n", "Be11dTOBe12p", "Be11dTOLi9a", "Be11He3TOC14g", 
      "Be11He3TOC13n", "Be11He3TOB13p", "Be11He3TOBe10a", "Be11pTOB12g", 
      "Be11tTOB14g", "Be11tTOB13n", "Be12aTOC16g", "Be12aTOC15n", 
      "Be12dTOB14g", "Be12dTOB13n", "Be12He3TOC15g", "Be12He3TOC14n", 
      "Be12He3TOB14p", "Be12He3TOBe11a", "Be12pTOB13g", "Be12pTOB12n", 
      "Be12tTOB15g", "Be12tTOB14n", "B8aTOC11p", "B8dTOC10g", 
      "B8He3TOC10p", "B8tTOC11g", "B8tTOC10n", "B8tTOB10p", "B8tTOBe7a", 
      "B10dTOC12g", "B10dTOC11n", "B10dTOB11p", "B10dTOaaa", 
      "B10He3TON13g", "B10He3TON12n", "B10He3TOC12p", "B10nTOBe10p", 
      "B10tTOC13g", "B10tTOC12n", "B10tTOB12p", "B10tTOBe9a", 
      "B11dTOC13g", "B11dTOC12n", "B11dTOB12p", "B11dTOBe9a", 
      "B11He3TON14g", "B11He3TON13n", "B11He3TOC13p", "B11He3TOB10a", 
      "B11tTOC14g", "B11tTOC13n", "B11tTOBe10a", "B12aTON16g", 
      "B12pTOC12n", "B12aTON15n", "B12dTOC14g", "B12dTOC13n", 
      "B12dTOB13p", "B12dTOBe10a", "B12He3TON15g", "B12He3TON14n", 
      "B12He3TOC14p", "B12He3TOB11a", "B12nTOB13g", "B12pTOC13g", 
      "B12tTOC15g", "B12tTOC14n", "B12tTOBe11a", "C9aTOO13g", 
      "C9dTOC10p", "C9nTOC10g", "C9tTON12g", "C9tTOC11p", "C9tTOB8a", 
      "C11dTON13g", "C11dTOC12p", "C11He3TOO14g", "C11He3TON13p", 
      "C11He3TOC10a", "C11tTON14g", "C11tTON13n", "C11tTOC13p", 
      "C11tTOB10a", "C12aTOO16g", "C12dTON14g", "C12dTOC13p", 
      "C12He3TOO15g", "C12He3TON14p", "C11aTOO15g", "C12He3TOC11a", 
      "C12nTOC13g", "C12pTON13g", "C12tTON15g", "C12tTON14n", 
      "C12tTOC14p", "C12tTOB11a", "C13aTOO17g", "C13dTON15g", 
      "C13dTON14n", "C13dTOC14p", "C13dTOB11a", "C13He3TOO16g", 
      "C13He3TOO15n", "C13He3TON15p", "C13He3TOC12a", "C13nTOC14g", 
      "C13pTON14g", "C13tTON16g", "C13tTON15n", "C13tTOC15p", 
      "C13tTOB12a", "C14dTON16g", "C14dTOB12a", "C14He3TOO17g", 
      "C14He3TOO16n", "C14He3TON16p", "C14He3TOC13a", "C14tTON17g", 
      "C14tTON16n", "C15aTOO19g", "C15aTOO18n", "C15nTOC16g", 
      "C15pTON16g", "C15pTON15n", "C15pTOB12a", "N12aTOO15p", 
      "N12nTON13g", "N12pTOO13g", "N13aTOF17g", "N13nTOC13p", 
      "N13pTOO14g", "N14nTON15g", "N14pTOO15g", "N15nTON16g", 
      "N15pTOO16g", "O14nTOO15g", "O14nTOC11a", "O15nTOO16g", 
      "O15nTON15p", "O15nTOC12a", "O17aTONe21g", "O17aTONe20n", 
      "O19aTONe23g", "O19aTONe22n", "O19nTOO20g", "O19pTOF20g", 
      "O19pTOF19n", "O19pTON16a", "F17aTONa21g", "F17aTONe20p", 
      "F17nTOF18g", "F18aTONa22g", "F18aTONe21p", "F18nTOF19g", 
      "F19aTONa23g", "F19aTONe22p", "F19nTOF20g", "F19pTONe20g", 
      "F19pTOO16a", "B8nTOLi6He3", "Li9pTOLi7t", "B8nTOBe7d", 
      "C9nTOBe7He3", "B10nTOaat", "Be10pTOaat", "Be11pTOBe9t", 
      "Be11pTOBe10d", "Be12pTOBe10t", "C9nTOB8d", "N13nTOC12d", 
      "B10aTOC12d", "O14nTOC12He3", "C15pTOC14d", "Ne18nTOO15a", 
      "Ne19nTOO16a", "Na20nTOF17a", "Ne18nTOF18p", "Ne19nTOF19p", 
      "Li7He3TOBe9p", "Li6tTOLi7d", "Li6He3TOBe7d", "Li7He3TOaad", 
      "Li8He3TOBe9d", "Li8He3TOaat", "Li9dTOLi8t", "Li9He3TOBe10d", 
      "Li9He3TOBe9t", "Be7tTOaad", "Be7tTOLi7He3", "Be9dTOaat", 
      "Be9tTOBe10d", "Be9He3TOB10d", "Be10He3TOB11d", "Be10He3TOB10t", 
      "B8dTOBe7He3", "B8tTOaaHe3", "B10pTOaaHe3", "B10tTOB11d", 
      "B10He3TOC11d", "B11tTOB13p", "B11He3TOC12d", "N12nTOC11d", 
      "C11tTOC12d", "C11tTOB11He3", "Be7He3TOppaa", "ddTOag", 
      "He3He3TOapp", "Li7aTOB11g", "Be7pTOB8g", "Be7aTOC11g", 
      "Be9pTOB10g", "Be9pTOaad", "Be9pTOLi6a", "Be9aTOC12n", 
      "B10pTOC11g", "B10pTOBe7a", "B11pTOC12g", "B11pTOaaa", 
      "B11aTON14n", "C13aTOO16n", "N15pTOC12a", "Li7tTOBe9n", 
      "B11nTOB12g", "C11nTOaaa", "Li7dTOaan", "dnTOtg", "ttTOann", 
      "He3nTOag", "He3tTOad", "He3tTOanp", "aanTOBe9g", "Li7tTOaann", 
      "Li7He3TOaanp", "Li8dTOLi9p", "Li8dTOLi7t", "Be7dTOaap", 
      "Be7tTOaanp", "Be7He3TOaapp", "C9aTON12p", "Li6nTOta", 
      "He3tTOLi6g", "anpTOLi6g", "Li6nTOLi7g", "Li6dTOLi7p", 
      "Li6dTOBe7n", "Li6aTOB10g", "Li7aTOB10n", "Li7nTOLi8g", 
      "Li7dTOLi8p", "Li8nTOLi9g", "Li8pTOaan", "Li8dTOBe9n", 
      "Be9nTOBe10g", "Be9pTOaapn", "B11pTOC11n", "Be10nTOBe11g", 
      "Be11nTOBe12g", "B8pTOC9g", "aaaTOC12gg", "C11pTON12g", 
      "B10aTON13n", "B11aTOC14p", "C11nTOC12g", "He6TOLi6Bm", 
      "Li8TOaaBm", "Li9TOBe9Bm", "Li9TOaanBm", "Be11TOB11Bm", 
      "Be12TOB12Bm", "B8TOaaBp", "B12TOC12Bm", "B13TOC13Bm", 
      "B14TOC14Bm", "B15TOC15Bm", "C9TOaapBp", "C10TOB10Bp", 
      "C11TOB11Bp", "C15TON15Bm", "C16TON16Bm", "N12TOC12Bp", 
      "N13TOC13Bp", "N16TOO16Bm", "N17TOO16nBm", "O13TON13Bp", 
      "O14TON14Bp", "O15TON15Bp", "O19TOF19Bm", "O20TOF20Bm", 
      "F17TOO17Bp", "F18TOO18Bp", "F20TONe20Bm", "Ne18TOF18Bp", 
      "Ne19TOF19Bp", "Ne23TONa23Bm", "Na20TONe20Bp", "Na21TONe21Bp", 
      "annTOHe6g", "O16nTOO17g", "N14nTOC14p", "O14nTON14p", 
      "O14aTONe18g", "C11aTON14p", "O14aTOF17p", "O17nTOC14a", 
      "F17nTON14a", "F18nTON15a", "C14dTON15n", "ppnTOdp", "C14nTOC15g", 
      "O16pTON13a", "Li8pTOBe9g", "B11aTON15g"};
        \end{lstlisting}
        }
        \item{In the same subsection, we redefine or add new functions to load the reaction network with the desired reaction rate increased to its $1\sigma$ upper limit:
        \begin{lstlisting}
  (* Include alternate definitions of the LoadRates function, 
  which loads in the reaction rates for the PRIMAT reaction network, 
  to accommodate the changes made to the code for the covariance estimation *)
  LoadRatesNoCovariance := (
      \[Tau]neutron = Mean\[Tau]neutron;
      LnTOp[Tv_] := 1/\[Tau]neutron*\[Lambda]nTOpI[Tv];
      LpTOn[Tv_] := 1/\[Tau]neutron*\[Lambda]pTOnI[Tv];
      LbarnTOp[Tv_] := LpTOn[Tv];
            
      ChosenName = "Placeholder";
      (* The list of reactions which are tabulated in external .dat file*)
      ListReactionsFile = TreatReactionLine /@ ReshapedTabulatedReactions; 
            
      (* The list of reactions defined analytically *)
      ExtraAnalyticReactions = DefineAnalyticRates;
            
      (* We concatenate the weak reactions, the tabulated rates and the analytic rates*)
      ListReactions = Join[{ReactionPEN}, ListReactionsFile, ExtraAnalyticReactions];
  );
        \end{lstlisting}
        \begin{lstlisting}
  LoadRatesCovariance[TweakedReactionNumber_] := (
      If[
          TweakedReactionNumber == 0,
          LoadRatesNoCovariance;,(*0 loads the central values*)
          (*If we're tweaking reaction 1 (n\[TwoWayRule]p), 
          include the uncertainty in the neutron lifetime (which dominates \
          the weak rate uncertainty).*)
          If[
              TweakedReactionNumber == 1, 
              \[Tau]neutron = Mean\[Tau]neutron + \[Sigma]\[Tau]neutron;, 
              \[Tau]neutron = Mean\[Tau]neutron;
          ];
            
          LnTOp[Tv_] := 1/\[Tau]neutron*\[Lambda]nTOpI[Tv];
          LpTOn[Tv_] := 1/\[Tau]neutron*\[Lambda]pTOnI[Tv];
          LbarnTOp[Tv_] := LpTOn[Tv];
            
          ChosenName = NamesArray[[TweakedReactionNumber]];
          (* The list of reactions which are tabulated in external .dat file*)
          ListReactionsFile = TreatReactionLine /@ ReshapedTabulatedReactions;
            
          (* The list of reactions defined analytically *)
          ExtraAnalyticReactions = DefineAnalyticRates;
            
          (* We concatenate the weak reactions, the tabulated rates and the analytic rates*)
          ListReactions = Join[{ReactionPEN}, ListReactionsFile, ExtraAnalyticReactions];
      ];
  );
        \end{lstlisting}
        \begin{lstlisting}
  LoadRates[NumberifCovariance_] := If[
      $LinearCovarianceEstimation, 
      LoadRatesCovariance[NumberifCovariance], 
      LoadRatesNoCovariance
  ]
        \end{lstlisting}
        \begin{lstlisting}
  LoadRates[0]; (*Input value other than zero directs PRIMAT to adjust the rate 
  of the reaction number (see NamesArray) to the upper 1sigma value*)
        \end{lstlisting}
        }       
    \end{enumerate}
    }
    \item{Finally, we tweak the \texttt{RunNumericalIntegrals} function to ensure the precomputed thermodynamics are loaded in and add an additional function to \texttt{Time integration of Cosmology and BBN} $\rightarrow$ \texttt{Gathering integrations on all periods in one function} to run \texttt{PRIMAT} with the functions defined for linear covariance estimation:
    \begin{lstlisting}
  (* Specify the precomputed thermodynamics from nudec_BSM should be loaded before running PRIMAT.  
  Also, include an additional function to allow the user to run PRIMAT with one reaction rate 
  adjusted for covariance estimation. *)
  RunNumericalIntegrals := (
      LoadNudecBSM;
        
      If[$RecomputeWeakRates, PreComputeWeakRates;
          \[Lambda]nTOpI = MyInterpolationRate[ToExpression[TabRatenp]];
          \[Lambda]pTOnI = MyInterpolationRate[ToExpression[TabRatepn]];
      ];
    \end{lstlisting}
    \qquad \ldots
    \begin{lstlisting}
      InterpolateResults;
  );
    \end{lstlisting}
    \begin{lstlisting}
  RunNumericalIntegralsLinEstimation[ReactionNumber_] := (
      LoadNudecBSM;

      (*Check to see if the weak rate file exists; if not, calculate weak rates*)
      If[$RecomputeWeakRates, PreComputeWeakRates;
          \[Lambda]nTOpI = MyInterpolationRate[ToExpression[TabRatenp]];
          \[Lambda]pTOnI = MyInterpolationRate[ToExpression[TabRatepn]];
      ];
        
      (* Build equations. Needed since rate are modified randomly by the f factor of each reaction*)
      LoadRates[ReactionNumber];
      DefineEquations;
        
      (* High temperature integration with only PEN reactions *)
      SolveValueHighTemperatures;
        
      (* Middle and Low temperature WITH nuclear reactions*)
      RunNumericalIntegralsNuclearReactions;
        
      InterpolateResults;
  );
    \end{lstlisting}
    Either of the above functions can be called to compute the abundances of light elements after BBN, depending on the user's needs.}
\end{enumerate}

\section{Fast Numerical Evaluation of Bose-Einstein and Fermi-Dirac Integrals}\label{app:integrals}

\setcounter{equation}{0}
\setcounter{figure}{0}
\setcounter{table}{0}
\renewcommand{\theequation}{D\arabic{equation}}
\renewcommand{\thefigure}{D\arabic{figure}}
\renewcommand{\thetable}{D\arabic{table}}

To solve the differential equations governing the evolution of the photon and neutrino temperatures prior to and during BBN, the energy density $\rho$, pressure $P$, and $d \rho / dT$ of all the fluids in both the electromagnetic and the neutrino sectors need to be evaluated at each time step. In \texttt{nudec\_BSM}, these quantities are evaluated using the default numerical quadrature method in \texttt{SciPy}; performing these numerical integrals takes up the majority of the time in each individual run of \texttt{nudec\_BSM}, and makes the code unsuitable for large parameter scans involving nonzero $\Delta N_{\nu}$. 

To improve the speed of the code, we adopt a method that was developed for numerical integration over a photon blackbody distribution in Ref.~\cite{Liu:2019bbm}: integrals over a Fermi-Dirac or Bose-Einstein distribution are rewritten as an infinite converging series, and the numerical value of the integral approximated by evaluating a sufficient number of leading terms in the series. This method of evaluating integrals leads to an order of magnitude speed-up in the \texttt{nudec\_BSM} code compared to the default numerical quadrature method. 

To obtain $\rho$, $P$, and $d\rho/dT$ for a Fermi-Dirac or a Bose-Einstein gas, we need to numerically evaluate the following integrals: 
\begin{alignat}{1}
    I_{3/2}^\pm (a, b) = \int_a^\infty d\xi \frac{(\xi^2 - a^2)^{3/2}}{e^{b\xi} \pm 1} \,, \qquad I_{1/2}^\pm (a, b) = \int_a^\infty d\xi \frac{\sqrt{\xi^2 - a^2}}{e^{b\xi} \pm 1} \,,
\end{alignat}
where the $+$($-$) sign is taken for a Fermi-Dirac (Bose-Einstein) gas. For a particle with internal degrees of freedom $g$, mass $m$, and temperature $T$, the relevant quantities are related to these integrals by
 \begin{alignat}{1}
     \rho_\pm &= \frac{g T^4}{2 \pi^2} \left[I^\pm_{3/2}(x, 1) + x^2 I^\pm_{1/2}(x, 1) \right] \,, \nonumber \\
     P_\pm &= \frac{g T^4}{6 \pi^2} I_{3/2}^\pm(x, 1) \,, \nonumber \\
     \frac{d \rho_\pm}{dT} &= -\frac{g T^3}{2 \pi^2} \left[ \frac{\partial I_{3/2}^\pm}{\partial b} (x, 1) + x^2 \frac{\partial I_{1/2}^\pm}{\partial b}(x, 1) \right] \,,
 \end{alignat}
 where $x \equiv m/T$. 

We begin by writing $I_{1/2}^\pm$ as a convergent infinite series of integrals that can be analytically performed. First, we rewrite
\begin{alignat}{1}
    \frac{1}{e^{b \xi} \pm 1} = e^{-b \xi} \sum_{k=0}^\infty (\mp 1)^k e^{-k b \xi} \,,\nonumber
\end{alignat}
so that the integral now becomes
\begin{alignat}{1}
    I_{1/2}^\pm (a, b) = \sum_{k=0}^\infty (\mp 1)^k \int_a^\infty d\xi \, \sqrt{\xi^2 - a^2} e^{-(k+1) b \xi} \,,
\end{alignat}
which can be performed in a computer algebra system to obtain
\begin{alignat}{1}
    I_{1/2}^\pm (a, b) = \sum_{k=0}^\infty (\mp 1)^k \frac{a}{b(k+1)} K_1(a b (k+1)) \,,
\end{alignat}
where $K_1$ is the modified Bessel function of the second kind. A similar exercise for $I_{3/2}^\pm$ leads to
\begin{alignat}{1}
    I_{3/2}^\pm (a, b) = \sum_{k=0}^\infty (\mp 1)^k \frac{3a^2}{b^2 (k+1)^2} K_2(a b (k+1)) \,.
\end{alignat}
The integrals that we need are therefore
\begin{alignat}{2}
    I_{1/2}^\pm (x, 1) &= \sum_{k=0}^\infty (\mp 1)^k \frac{x}{k+1} K_1((k+1)x) \,, \qquad I_{3/2}^\pm (x, 1) &&= \sum_{k=0}^\infty (\mp 1)^k \frac{3x^2}{(k+1)^2} K_2((k+1)x) \,, \nonumber \\
    \frac{\partial I_{1/2}^\pm}{\partial b}(x, 1) &= - \sum_{k=0}^\infty (\mp 1)^k x^2 K_2((k+1)x) \,, \quad\ \  \frac{\partial I_{3/2}^\pm}{\partial b}(x, 1) &&= - \sum_{k=0}^\infty (\mp 1)^k \frac{3x^3}{k+1} K_3((k+1) x) \,.
\end{alignat}
These series converge very quickly for large $x$: for $x > 0.01$, we find that evaluating the first 20 terms in the series is sufficient to obtain a result that is within 0.1\% of the result from the default \texttt{SciPy} numerical quadrature method. For $x < 0.01$, we simply set $x = 0$ (i.e., taking the massless limit) and evaluating the integral analytically, which gives a result that is also within 0.1\% of the result from the default \texttt{SciPy} numerical quadrature method. For completeness, $\rho$, $P$, and $d \rho / dT$ for a massless Bose-Einstein gas with internal degrees of freedom $g$ and temperature $T$ is 
\begin{alignat}{1}
    \rho_- = \frac{g T^4}{2 \pi^2} \frac{\pi^4}{15} \,, \qquad P_- = \frac{1}{3}\rho \,, \qquad \frac{d \rho_-}{dT} = \frac{4 \rho}{T} \,.
\end{alignat}
For a massless Fermi-Dirac gas, $\rho_+ = (7/8) \rho_-$ for the same $g$ and $T$, with $P_+ = \rho_+ / 3$ and $d \rho_+ / dT = 4 \rho_+ / T$ once again.

\end{document}